\documentclass[useAMS]{mn2e}
\usepackage{graphicx}

\def\lesssim{\,\lower2truept\hbox{${<\atop\hbox{\raise4truept\hbox{$\sim$}}}$}\,}
\def\gtrsim{\,\lower2truept\hbox{${>\atop\hbox{\raise4truept\hbox{$\sim$}}}$}\,}
    
    \def\smallskip{\vskip 6pt}
    
    \def\M12{${\rm M_rmn{12}}$}
    
\newcommand{\lsim}{\mathrel{\rlap{\raise -.3ex\hbox{${\scriptstyle\sim}$}}%
                   \raise .6ex\hbox{${\scriptstyle <}$}}}%
\newcommand{\gsim}{\mathrel{\rlap{\raise -.3ex\hbox{${\scriptstyle\sim}$}}%
                   \raise .6ex\hbox{${\scriptstyle >}$}}}%

\title[Star forming galaxies in the MIR]{The evolution of actively star forming galaxies in the mid infrared.}
\author[Vega et al.]{O. Vega$^{1}$, L. Silva$^{2}$, P. Panuzzo$^{3}$, A. Bressan$^{3}$, G. L. Granato$^{3}$, \& M. Chavez$^{1}$ \\
$^{1}$INAOE, Luis Enrique Erro 1, 72840, Tonantzintla, Puebla, Mexico \\
$^{2}$INAF-Trieste, Via Tiepolo 11, I-34131 Trieste, Italy \\
$^{3}$INAF-Padova, Vicolo dell'Osservatorio 5, I-35122 Padova, Italy}

\begin{document}

\date{Accepted. Received}
\pagerange{\pageref{firstpage}--\pageref{lastpage}} \pubyear{2004}
\maketitle

\label{firstpage}

\begin{abstract}
In this paper we analyze the evolution of actively star forming
galaxies in the mid-infrared (MIR). This spectral region,
characterized by continuum emission by hot dust and  by the
presence of strong emission features generally ascribed to
polycyclic aromatic hydrocarbon (PAH) molecules, is the most
strongly affected by the heating processes associated with star
formation and/or active galactic nuclei (AGN). Following the
detailed observational characterization of galaxies in the MIR by
ISO, we have updated the modelling of this spectral region in our
spectro-photometric model GRASIL (Silva et al. 1998). In the
diffuse component we have updated the treatment of PAHs according
to the model by Li \& Draine (2001). As for the dense phase of the
ISM associated with the star forming regions, the molecular
clouds, we strongly decrease the abundance of PAHs as compared to
that in the cirrus, basing on the observational evidences of the
lack or weakness of PAH bands close to the newly formed stars,
possibly due to the destruction of the molecules in strong UV
fields. The robustness of the model is checked by fitting near
infrared to radio broad band spectra and the corresponding
detailed MIR spectra of a large sample of galaxies (Lu et al.
2003), at once. With this model, we have analyzed the larger
sample of actively star forming galaxies by Dale et al. (2000). We
show that the observed trends of galaxies in the ISO-IRAS-Radio
color-color plots can be interpreted in terms of different
evolutionary phases of star formation activity, and the consequent
different dominance in the spectral energy distribution (SED) of
the diffuse or dense phase of the ISM. We find that the observed
colors indicate a surprising homogeneity of the starburst
phenomenon, allowing only a limited variation of the most
important physical parameters, such as the optical depth of the
molecular clouds, the escape timescale of young stars from their
formation sites, and the gas consumption timescale. In this paper
we do not attempt to reproduce the far infrared (FIR) coolest
region in the color-color plots, since we concentrate on models
meant to reproduce active star forming galaxies, but we discuss
possible requirements of a more complex modelling for the coldest
objects.
\end{abstract}

%%%%%%%%%%%%%%%%%%%%%%%%%%%%%%%%%%%%%%%%%%%%%%%%%

\section{Introduction}

With the advent of the new generation of infrared space telescopes
it has become possible to study in great detail the mid-infrared
(MIR) spectral region of galaxies, that had been previously
sampled only with  broadband filters by IRAS. Besides the hot
continuum emission, an important component observed in the MIR
spectra of many dusty galactic and extragalactic objects is
constituted by strong emission features at 3.3, 6.2, 7.7, 8.6,
11.3 and 12.7~$\mu$m. These features are most commonly ascribed to
aromatic C-C and C-H vibrations in large planar Polycyclic
Aromatic Hydrocarbons (PAH) molecules, with size $\sim$ 10
\rm{\AA}
 and containing $\sim 50-100$ C atoms (Leger \& Puget
1984; Allamandola et al. 1985; Puget \& Leger 1989; Tielens et al.
1999).

The environmental conditions that may contribute to the MIR
continuum emission are quite well understood in terms of a strong
radiation field that heats dust at the required temperatures. Such
regions are mainly located in the vicinity of hot stars, in the
AGN torii and in dense circumstellar envelopes of cool evolved
stars. Conversely our knowledge of the origin of PAH emission is
still controversial, in spite of its importance for an accurate
modelling of the MIR spectral region.

Detailed observations with ISO have allowed deep investigations of the properties and variations of
PAH features in different Galactic environments  - post-AGB stars, planetary nebulae, reflection
nebulae, young stellar objects, HII regions and photo-dissociation regions (PDR), diffuse
interstellar medium -, as well as in different galaxies (e.g. Cesarsky et al. 1996; Verstraete et
al. 1996; Beintema et al. 1996; Boulanger et al. 1996; Mattila et al. 1996; Peeters, Spoon, \&
Tielens 2004a; see the review by Peeters et al. 2004b and references therein). The aromatic
emission features in most galaxies have been found to be very similar to those in Galactic star
forming regions and in the cirrus (Metcalfe et al. 1996; Vigroux et al. 1996; Genzel et al. 1998;
Lutz et al. 1998; Rigopoulou et al. 1999; Tielens 1999; Helou et al. 2000; Forster Schreiber et al.
2003; Lu et al. 2003). However MIR spectra with different appearances have been observed in
peculiar environments, with particularly strong radiation fields (e.g. in AGNs, Genzel et al. 1998;
Lutz et al. 1998). This has  led to the use of  PAH emission as a tracer of the star formation
process (Roussel et al. 2001, Forster Schreiber et al. 2004), and its lack as indicative of an AGN
dominated emission mechanism (Genzel et al. 1998; Clavel et al. 2000; Laurent et al. 2000).

More recently, a comprehensive analysis of the MIR to FIR spectra
of Galactic star forming regions, normal galaxies and starbursts
suggests that PAHs are a better tracer of B stars rather than of
massive O stars (Peeters et al. 2004a, Boselli, Lequeux, \&
Gavazzi 2004, Madden 2005). This has been found also by
Tacconi-Garman et al. (2005) by direct high resolution 3 $\mu$m
imaging of two nearby starbursts (NGC253 and NGC1808): they find
no spatial coincidence between the detailed distribution of PAH
emission and the sites of the most recent star formation.

Several recent works have also shown that the intensity of the PAH
bands is systematically lower in very low metallicity galaxies,
such as Blue Compact Dwarfs and dwarf galaxies in general, and
some irregulars (e.g. Madden 2000; Galliano et al. 2003, 2005;
Boselli et al. 2004; Dale et al. 2005; Engelbracht et al. 2005).
This has been ascribed either to a lower intrinsic PAH abundance
or to a higher destruction rate of these molecules due to a strong
UV field (see e.g. the review by Madden 2005).

In summary there is mounting evidence that the use of the MIR region and in particular of the PAH
features as tracers of star formation in galaxies is not straightforward.

In order to correctly interpret the spectra of galaxies, a
detailed modelling that includes the complex interaction between
the evolution of stellar populations and dust is needed. In fact
while the new observations have allowed us to better define the
characterization of PAHs in the diffuse medium (e.g. Li \& Draine
2001; Zubko, Dwek, \& Arendt 2004), most current
spectro-photometric models adopt a pure phenomenological approach
(e.g. Dale et al.\ 2001). A recent model that combines radiation
transfer within the star forming regions with the evolution of the
powering star clusters is presented by Dopita et al. (2005). This
model, though being the most advanced in the specific treatment of
the star forming regions, lacks the diffuse component and cannot
be safely compared with galaxies as a whole. On the other hand the
PAH treatment in our code GRASIL\footnote{ Available at
http://adlibitum.oat.ts.astro.it/silva/default.html or
http://webpd.astro.it/granato/grasil.html} (Silva et al. 1998),
that takes into account the entire temporal evolution of several
subcomponents, was still based on pre-ISO data.

For this reason we present here an updated version of GRASIL
including the new developments for the PAH emission, and perform
an analysis of the evolution of {\it normal metallicity} (i.e. not
dwarf galaxies whose MIR SEDs are observed to behave differently
as recalled above) actively star forming galaxies in the mid
infrared.

The structure of the paper is the following. In
Section~\ref{sec:grapah} we summarize the inclusion of the new
theoretical model of PAHs by Li \& Draine (2001, LD01 hereafter).
There we make the important assumption, supported by observations,
that the PAH abundance within our (star-forming) molecular clouds
(MC) is significantly depleted with respect to that in the cirrus
component. In Section~\ref{sec:callu} we perform a detailed
comparison of the new model with NIR to radio SEDs of star forming
galaxies taken from the Lu et al. (2003) sample. We show that,
together with the wide NIR-Radio broad band fluxes, also the
MIR-ISO narrow bands, dominated by PAH emission, are very well
reproduced. Since these galaxies span a wide range of FIR
luminosities, we conclude that a proper combination of a cirrus
component with PAH emission suited for the diffuse ISM in our
Galaxy (where the LD01 model is calibrated), and a MC component
with strongly depleted PAHs, can be safely adopted in general for
star forming galaxies. With this new model we explore the
MIR-FIR-radio colors of a larger sample of galaxies taken from
Dale et al. (2000) for which only broad-band ISO, IRAS and Radio
fluxes are available (Section~\ref{sec:colcol}). This comparison
allows us to draw a physical interpretation of the observed data,
in terms of the evolution of the starburst phenomenon. The MID-FIR
and Radio data constitute a unique set of constraints indicating a
strong homogeneity of the main physical parameters. Our
conclusions are summarized in Section~\ref{sec:conc}.

\section{PAH modelling in GRASIL}
\label{sec:grapah}

In this Section we provide a short summary of the properties of GRASIL. We defer the reader to
Silva et al. (1998) and Silva (1999) for further details.

GRASIL calculates the UV to radio SED of galaxies by taking into account a detailed treatment of
dust reprocessing of stellar radiation. A particular important point is that the coupling between
stars and dust is considered, by letting stars form inside optically thick molecular clouds (MCs)
and then gradually escape as they grow old. Older stellar generations are distributed within the
diffuse ISM (cirrus). This modelling, with a realistic 3-D treatment of extinction and IR emission,
allows quantitative physical interpretations of multi-wavelength observations as well as
predictions (e.g. Silva et al. 1998; Granato et al. 2000; Panuzzo et al. 2003, 2004).

In addition to the complex geometry, the SEDs depend on the dust
intrinsic properties (chemical composition, size distribution,
shape, see the recent reviews by Draine 2003 and Dwek 2004). These
properties depend on the different environments in which dust
grains form or evolve. Most of the available dust models are
suited for the Galactic cirrus, for which the largest data set is
available. The dust model we use in GRASIL is made of graphite and
silicate spherical dust grains (Draine \& Lee 1984 and Laor \&
Draine 1993), with a power-law size distribution that extends to
include also the very small grains transiently heated by single
photons, and by PAH molecules.  The dust heating and emission are
adequately computed for each component.

\subsection{PAH emission in the cirrus component}

In Silva et al. (1998) we took into account five PAH emission bands, at 3.3, 6.2, 7.7, 8.6 and
11.3~$\mu$m. We adopted the optical-UV absorption cross-section by Leger, D'Hendecourt \&
Defourneau (1989), obtained by laboratory measurements for different PAH mixtures. We followed
Leger et al. (1989) for the heat capacity as a function of temperature and number of C and H atoms
in the molecules. We adopted a population of PAHs with a continuous power-law distribution,
$dn/dN_\rmn{C} \propto N_\rmn{C}^{-2.25}$, with $N_\rmn{C}$, the number of Carbon atoms in the
molecules, extending from 20 to 280 atoms. This range corresponds to a size $a=4$ to $15$ \rm{\AA},
with $a=0.9 \, \sqrt{N_\rmn{C}}$ (valid for the catacondensed PAHs, the most compact and stable
configuration). We took into account the possibility of partial hydrogenation of the molecules, as
suggested by observations, and fixed the hydro-coverage (the ratio of the number of H atoms over
the number of available sites) so to reproduce the observed ratio of C-H and C-C bands in the
Galactic cirrus. The abundance of carbon atoms locked in PAHs in our dust model is set to 18 ppM in
the cirrus, and was 10 times lower in the MCs to account for PAH destruction in strong UV fields
(see Section~\ref{susec:mc}).

The PAH emissivity is computed following mainly Xu \& De Zotti
(1989). Once emitted, PAH bands are absorbed by MCs and/or cirrus
dust before emerging from the model galaxy. This treatment
provides a rigorous computation of the total energy emitted in
these features.

ISO data have allowed us to get a deeper knowledge of PAH
phenomenology, as summarized in the Introduction. In particular,
new PAH bands have been detected, their profiles studied in
different environments and galaxies, thus yielding several
constraints for their modelling. Nevertheless, the properties of
astrophysical PAHs are still rather uncertain, so that, as in the
case of dust models in general, their modelling still require
ad-hoc settings of parameters to best reproduce the observations.

Li \& Draine (2001) presented a quantitative model for dust in the diffuse ISM, including PAHs.
Basing on laboratory studies of these molecules, and by comparing with observations, they have
provided ``astronomical'' cross sections for PAHs, reproduced with Drude profiles. We have updated
our model for PAHs in GRASIL by adopting the main features of the LD01 model: in addition to the
3.3, 6.2, 7.7, 8.6, 11.3~$\mu$m bands, we now include also the 11.9, 12.7, 16.4, 18.3, 21.2, and
23.1~$\mu$m bands. Absorption and emission from PAHs are computed as described above, but adopting
the FWHM and emission cross sections by LD01. As stressed by the authors, available laboratory data
on PAHs differ by large factors among different groups, and also between experimental results and
theoretical calculations. Therefore, LD01 modified laboratory data for the emission cross sections
in order to reproduce observations (the Galactic cirrus emission).

The LD01 model for PAHs above $14 \mu$m is based on laboratory
data from Moutou et al. (1996), since galaxy spectroscopy was not
available. Very recently, some works have reported Spitzer spectra
that probe this poorly known region (e.g. Armus et al. 2004;
Peeters et al. 2004c; Smith et al. 2004; Weedman et al. 2005). In
some cases, several new PAH features have been observed between
$\sim 17$ and $22 \mu$m, but not coincident with the 18.3, 21.2,
and 23.1~$\mu$m bands of the model. Instead, in Smith et al.
(2004), no PAH band has been observed above $\sim 18 \mu$m in NGC
7331. Note anyway that these uncertainties are not expected to
affect our results. Indeed we have estimated the contribution of
the 18.3, 21.2, and 23.1~$\mu$m bands to the 25 $\mu$m IRAS filter
as compared to the continuum, for the models fitting the SEDs
described in Section \ref{sec:callu}. This estimate refers to the
worst case, in which no PAH features are present in galaxies above
$\sim 18 \mu$m. In most cases the contribution is less than 10\%,
only in a couple of cases it is $\lesssim 30$\%, i.e. we are
always within the IRAS observational errors. We also note that the
LD01 model fits very well the DIRBE data for dust emission in our
Galaxy (figures 8 and 10 in LD01), with one of the DIRBE filters
dominated by these three bands. It is conceivable that in a few
years it will become possible to have enough observational
constraints to better define a reference model for astrophysical
PAHs at long wavelengths.

\begin{table}
\centering \caption{Properties of the Lu et al. (2003) sub-sample. The galaxies are ordered by
increasing log($f_\nu (60 \mu \rmn{m}) /f_\nu (100 \mu \rmn{m})$). Column (3) displays the fraction
of the LW2 flux from the cirrus component over the total LW2 flux of the best fit model. Column (4)
shows the main contributor to the MIR SED of the best fit, "c" indicates that the MIR SED is
"cirrus emission dominated", "m" indicates that it is "molecular cloud emission dominated", and
"mix" that the MIR SED is a mixed of these two contributions, see Fig. \ref{fig:pahcalnew}. We
classify the fit as "c" if the ratio $F_\rmn{LW2}\rmn{(cirrus)}/F_\rmn{LW2}\rmn{(total)}$ is $\ge
2/3$, as "mix" if $1/3 \le F_\rmn{LW2}\rmn{(cirrus)}/F_\rmn{LW2}\rmn{(total)}< 2/3$ and as "m" if
$F_\rmn{LW2}\rmn{(cirrus)}/F_\rmn{LW2}\rmn{(total)}< 1/3$. }
  \label{tablu}
\begin{tabular}{|lcccccc|} \hline
  \hline
 Galaxy  &   $\log\left(\frac{f_\nu (60 \mu \rmn{m})}{f_\nu (100 \mu \rmn{m})}\right)$ &$\frac{F_\rmn{LW2}\rmn{(cirrus)}}{F_\rmn{LW2}\rmn{(total)}}$&  PAHs\\
         &                        &   &fit\\
  \hline
NGC6286 &  -0.44 &0.851&c\\
NGC3583 & -0.42  &0.744&c\\
IC3908  &  -0.33 &0.768&c\\
UGC2238 & -0.30  &0.785&c\\
NGC7771 & -0.30  &0.694&c\\
NGC5713 &  -0.24 &0.543&mix\\
NGC2388 &  -0.17 &0.461&mix\\
NGC4102 & -0.16  &0.460&mix\\
NGC3885 & -0.15  &0.541&mix\\
NGC1482 &  -0.14 &0.471&mix\\
NGC4691 &  -0.14 &0.384&mix\\
NGC1022 &  -0.13  &0.253&m\\
NGC1222 &  -0.07 &0.211&m\\
NGC4194 & -0.01  &0.062&m\\
\hline
\end{tabular}
\end{table}

\subsection{PAH emission in the MC component}
\label{susec:mc}

While the abundance of PAHs in the cirrus environment is
relatively constrained, the situation is much less clear for the
regions of intense star formation, characterized by higher
particle densities and/or larger UV radiation fields. Abergel et
al. (1994) already noticed a lack of PAH emission in dense
molecular clouds revealed by  $^\rmn{13}$CO. Marty et al. (1994)
advanced a possible explanation suggesting that in these clouds
PAH molecules could be depleted by accretion onto dust grains
and/or formation of organometallic complexes (see also Serra et
al. 1992). Concerning the effects of strong UV radiation field,
Allain, Leach, \& Sedlmayr (1996) showed that PAH molecules which
are already either ionized or even partially dehydrogenated, have
photo-dissociation rates far in excess of those of the neutral
species.

Only after ISO data were available, it was possible to build a
clearer and more direct picture of PAH emission, revealing a
general lack of such features in regions associated with the star
formation processes. Observations in our Galaxy show  a clear
decreasing of PAH emission  near hot stars, or when passing inward
from PDRs to HII regions (Boulanger et al. 2000; Contursi et al.
2000). Hony et al. (2001) report observations of a high degree of
dehydrogenation in star forming regions. Peeters, Spoon, \&
Tielens (2004a) show a lack of the PAH features in spectra of
ultra compact HII regions, and in that of heavily obscured
starbursts, both characterized by high dust optical depths. These
latter observations are particularly relevant because they raise
the question whether such a lack is due to a strong attenuation
accompanied by dilution of the continuum, or to a real destruction
of PAH molecules in such environments. To clarify this point we
performed several tests with models of molecular clouds with
GRASIL, varying the optical depth, the abundance of PAHs and the
sublimation temperature of the dust (the latter in order to allow
dilution by hot dust). It resulted that the only viable way to
suppress PAH emission in our MC models was to deplete their
abundance by a large factor ($\simeq$ 1000 with respect to that of
the cirrus in order to reproduce the MIR observations described
below). Thus we will assume in the following that the abundance of
PAHs in our MCs is fixed to $10^{-3}$ times that in the cirrus.
This is even more evident if one thinks that the lack of the
3.3~$\mu$m PAH emission near very hot stars observed by
Tacconi-Garman et al. (2005) is difficult to ascribe to dilution
by the continuum.

The necessity of such a strong damping has been considered also by
Dale et al. (2001). In their phenomenological IR SEDs, they
adopted an analytical damping factor for PAHs, function of the
intensity of the radiation field. From their treatment one sees
that when the radiation intensity is larger than a few thousands
time the local interstellar radiation field, the PAHs are almost
completely depleted.

Finally let us comment that observations indicate that PAH
emission in star forming environments arise mainly from PDR
regions. These regions are implicitly accounted for in our
treatment, because it accounts for absorbtion of dissociating
photons in the interior of the molecular clouds and in the
surrounding diffuse medium. In principle our code allows us to
consider a distribution of MCs with different optical depths
(Silva 1999), in particular we could include a class of low
optical depth molecular clouds where PAHs are not depleted. While
this would allow us to explicitly model the PDR regions (and will
be the subject of a forthcoming paper), this would introduce
another set of free parameters that appear not to be relevant for
the actively star forming galaxies we compare with below, as
illustrated in the following sections.

\begin{figure*}
\centering
\includegraphics[angle=0,width=8.8truecm]{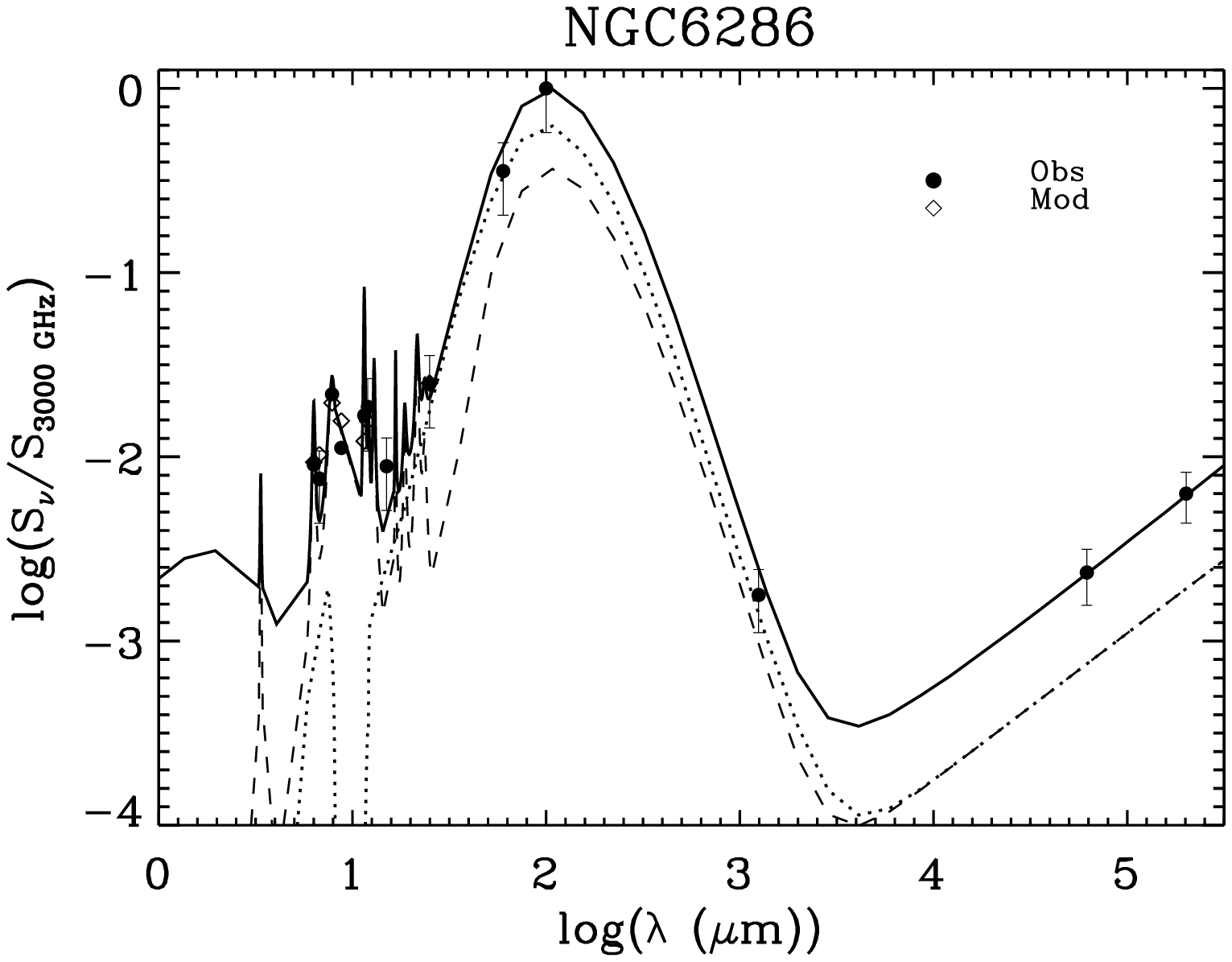}
\includegraphics[angle=0,width=8.8truecm]{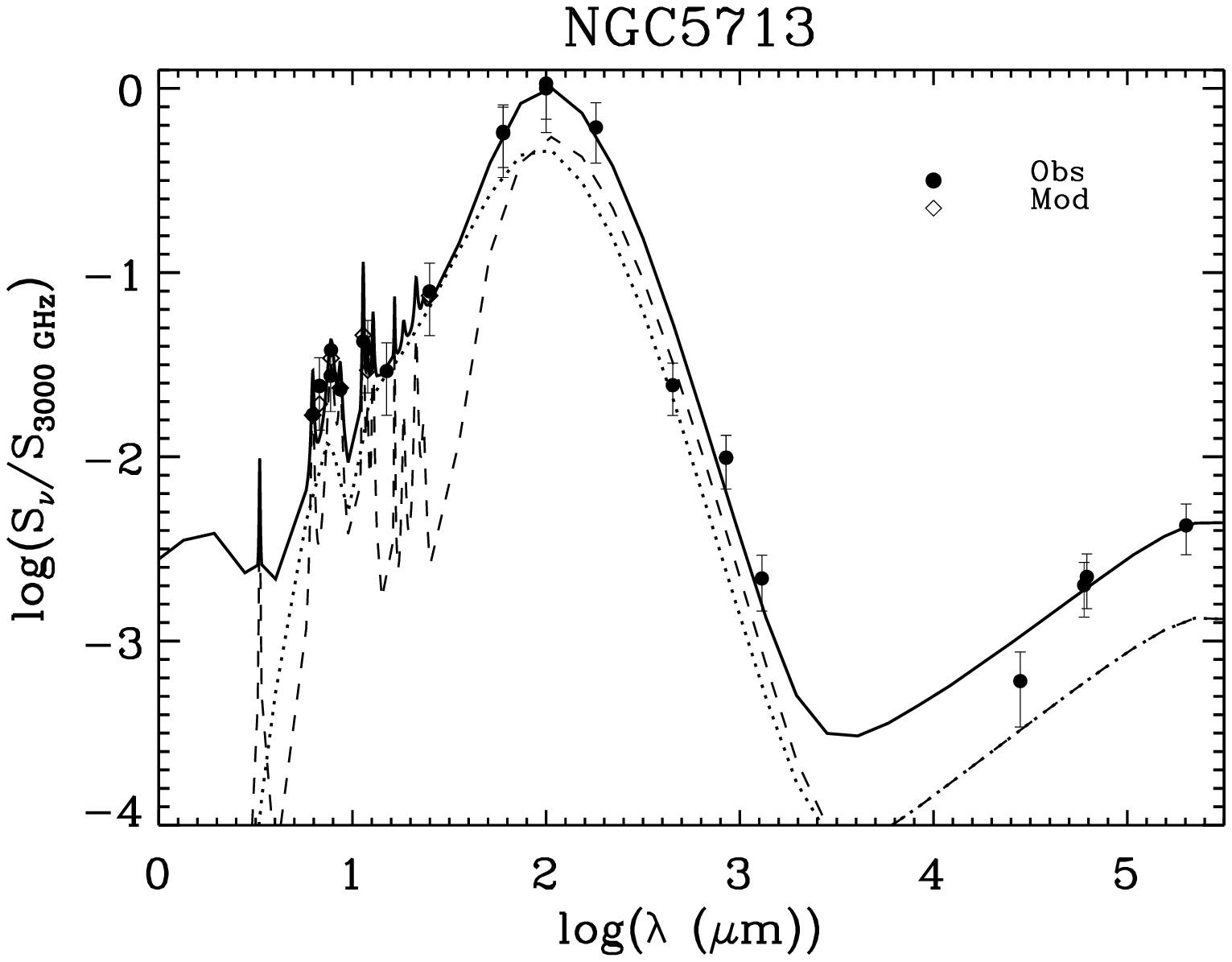}
\includegraphics[angle=0,width=8.8truecm]{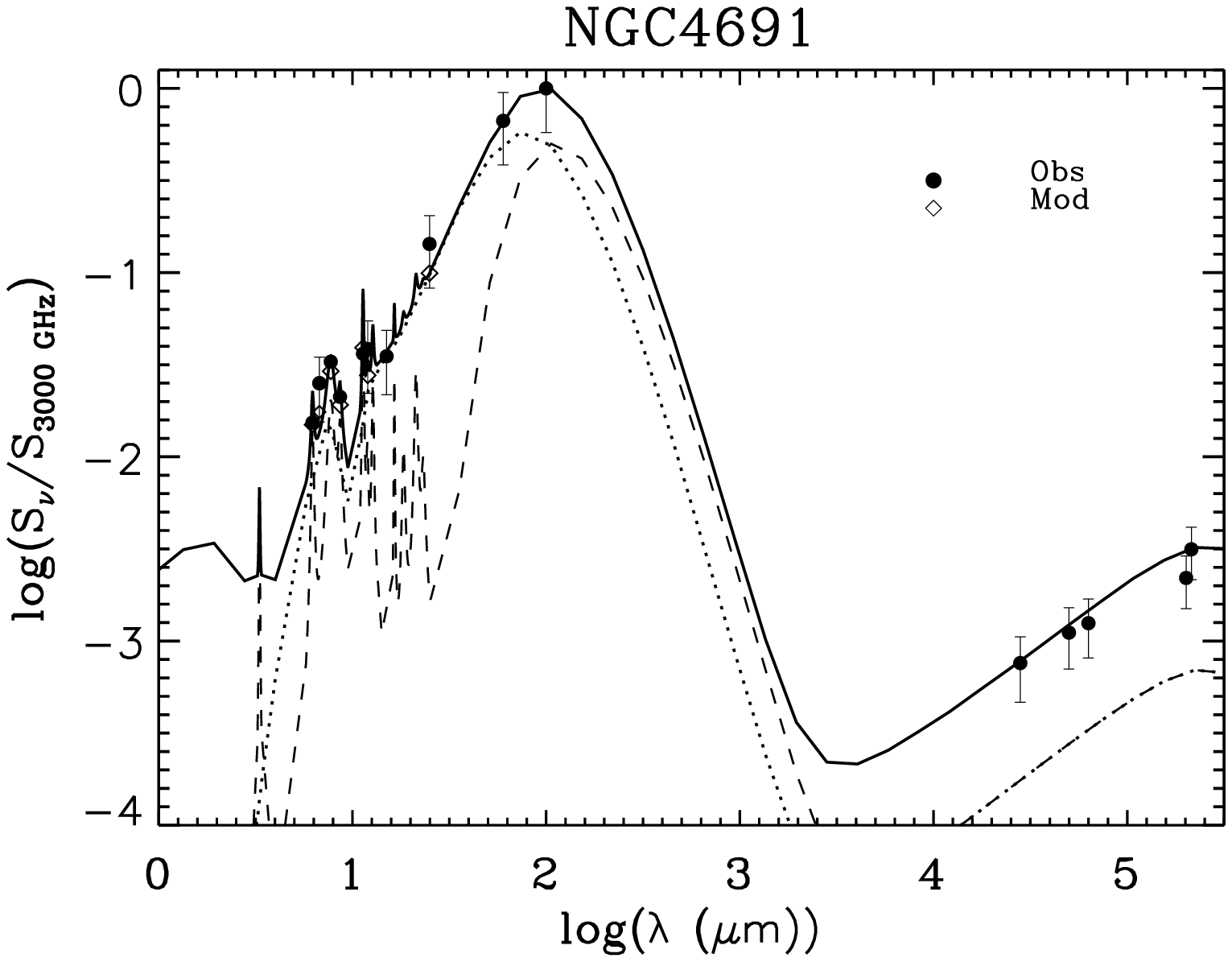}
\includegraphics[angle=0,width=8.8truecm]{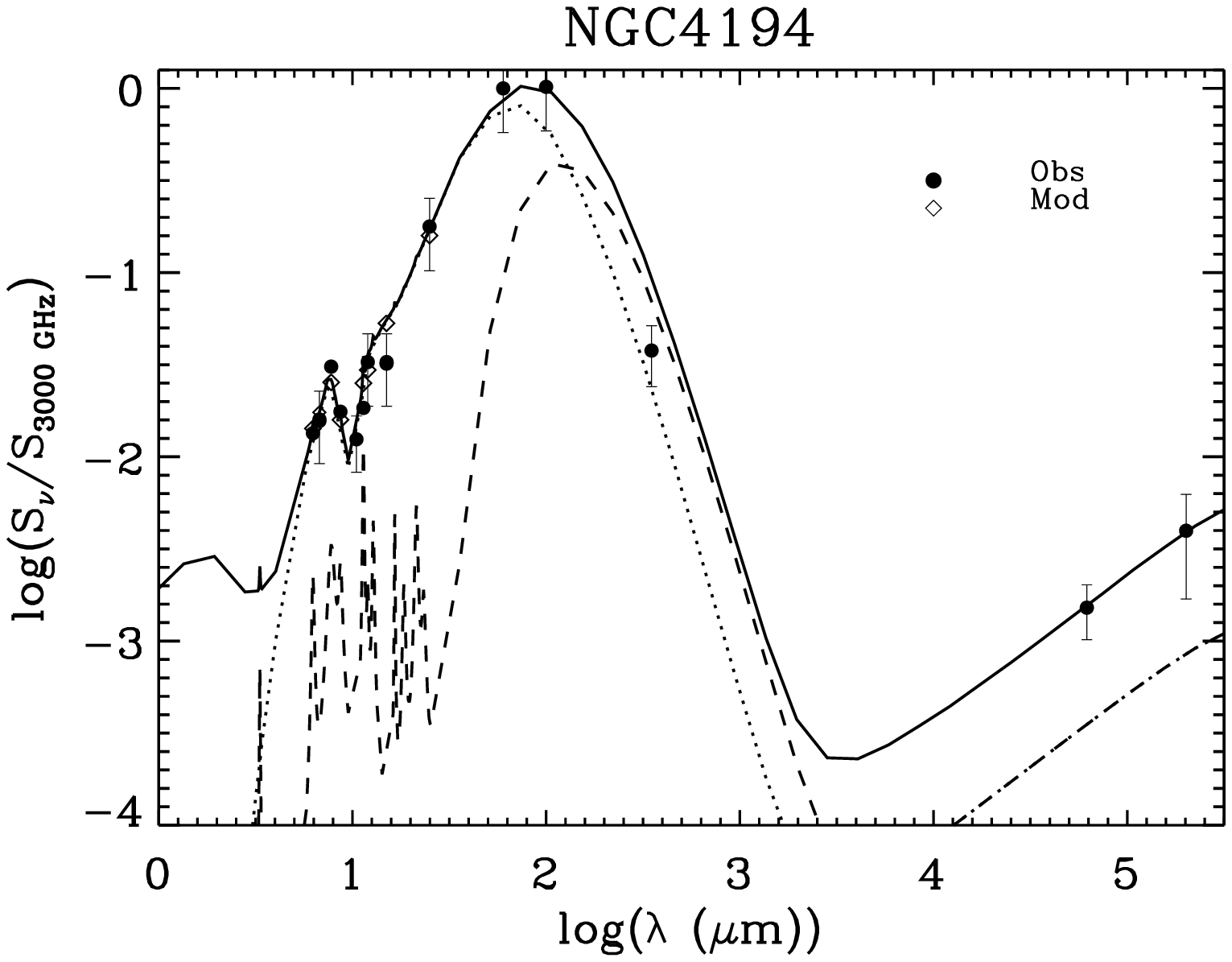}
\caption{Comparison between the observed and model SEDs for four
representative galaxies of our sample. The solid line is the total
emission, the dashed line is the contribution from the diffuse
medium, and the dotted line is the contribution from the molecular
clouds. Filled circles are the observational data by Lu et al.
(2003), Bendo et al. (2002), Dale et al. (2000), Dunne et al.
(2000), Machalski $\&$ Condon (1999), Chini et al. (1996, 1995),
Condon et al. (1995, 1990), Niklas et al (1995), Carico et al
(1992), and Israel et al. (1983). Open rombs are for the model in
the following order: 6.2$\mu$m, 6.75$\mu$m (ISO LW2), 7.7$\mu$m,
8.6$\mu$m, 11.3$\mu$m, 12$\mu$m (IRAS), 15$\mu$m (ISO LW3) and
25$\mu$m (IRAS).}
 \label{fig:totalsed}
\end{figure*}

\begin{figure*}
\centering
\includegraphics[angle=0,width=8truecm]{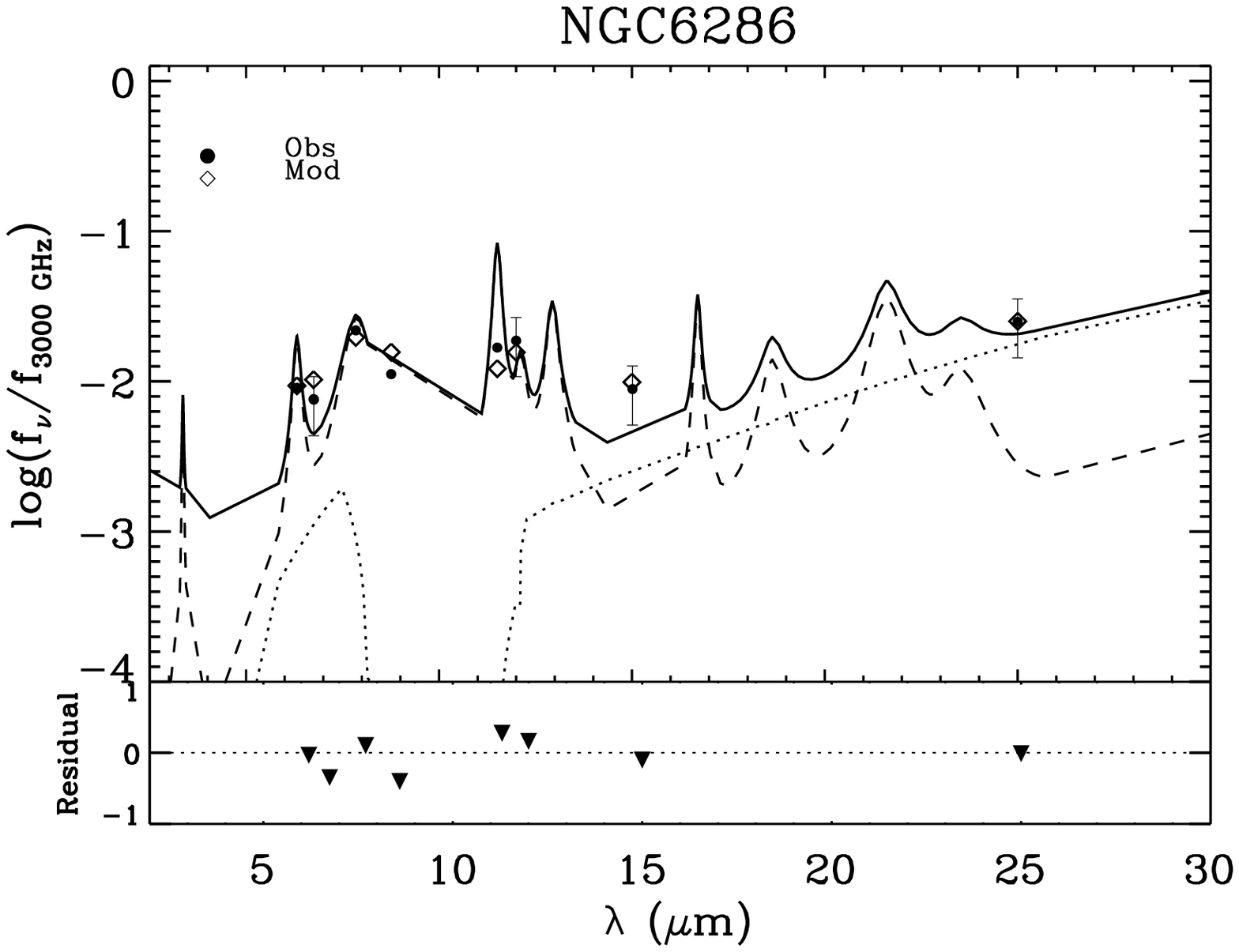}
\includegraphics[angle=0,width=8truecm]{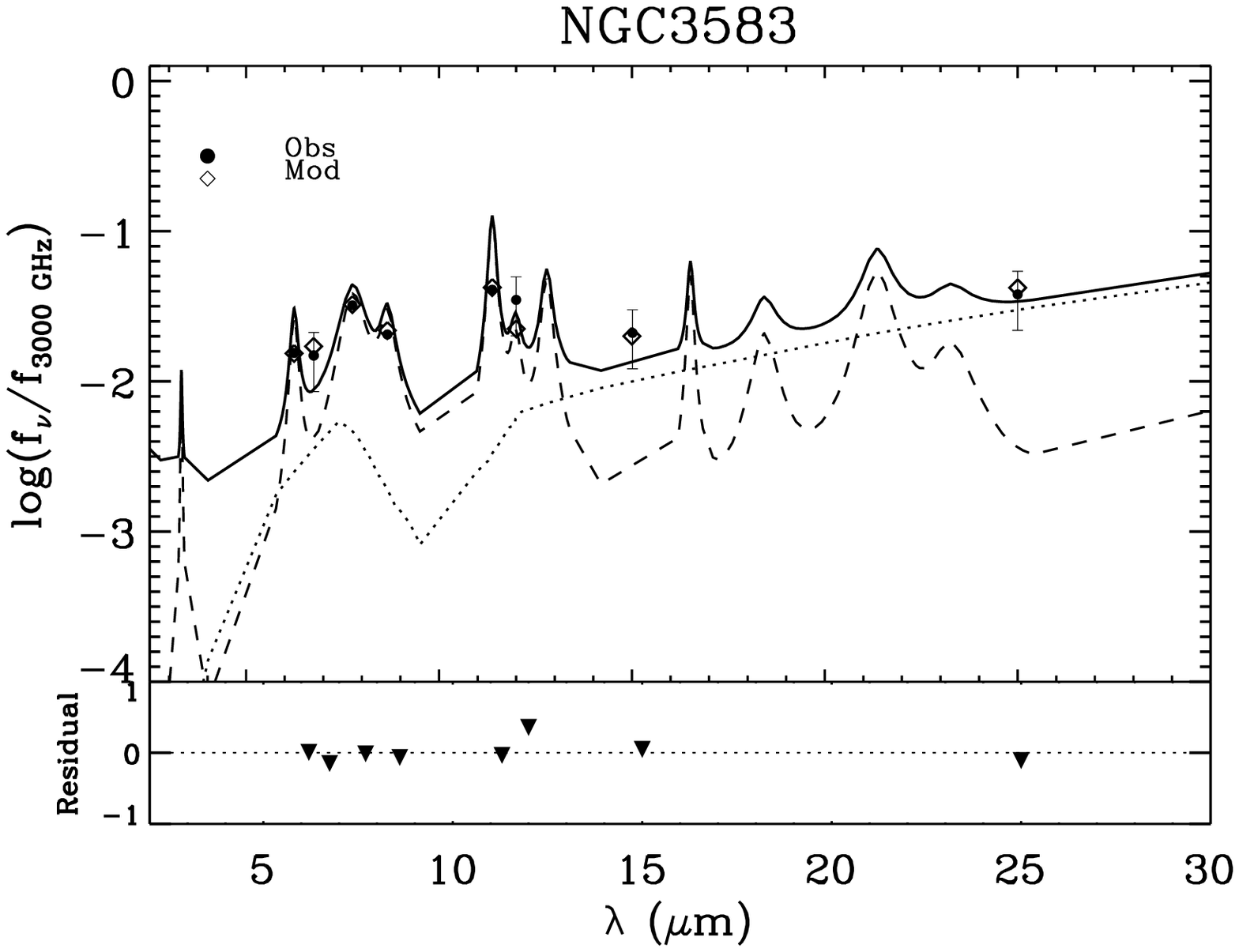}
\includegraphics[angle=0,width=8truecm]{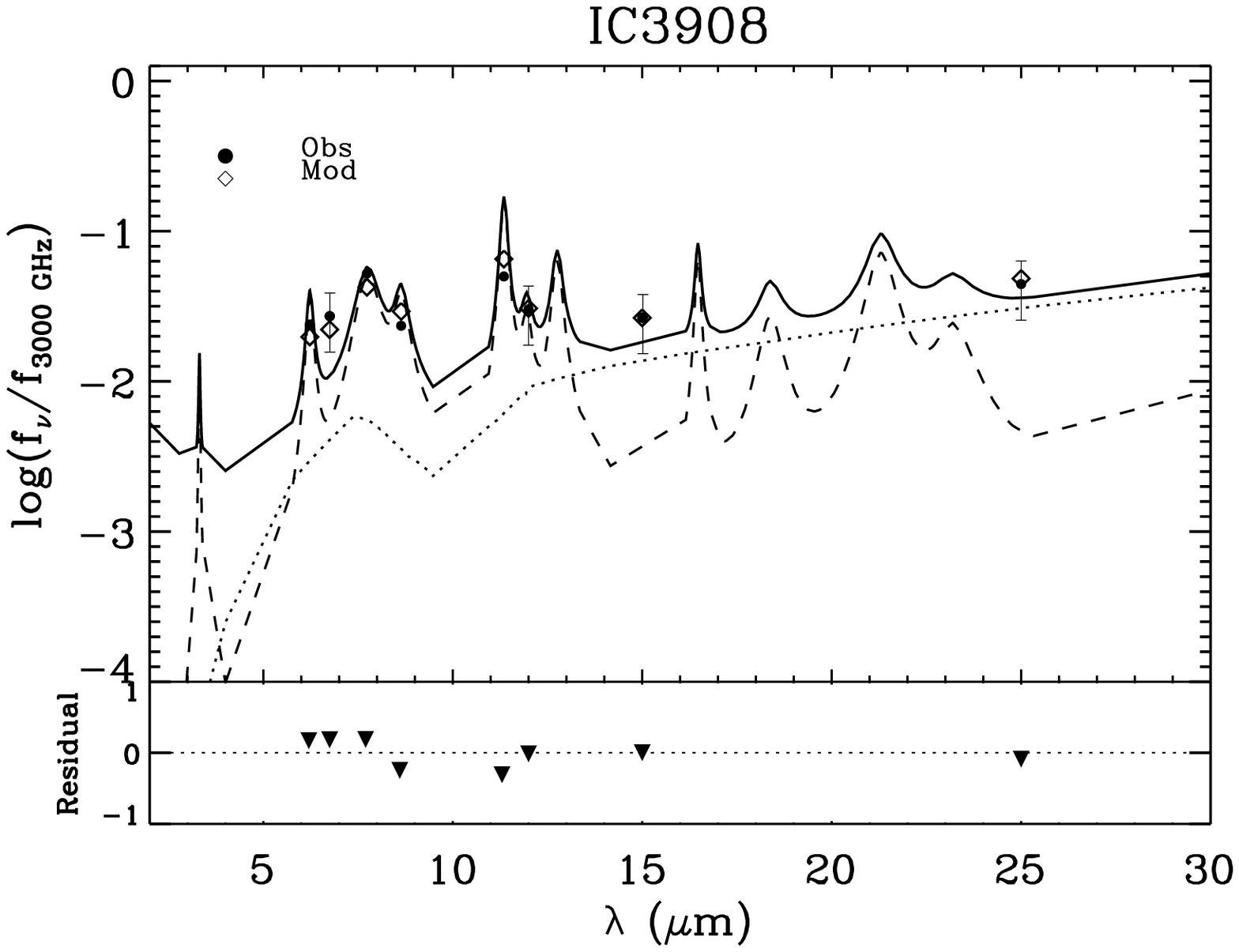}
\includegraphics[angle=0,width=8truecm]{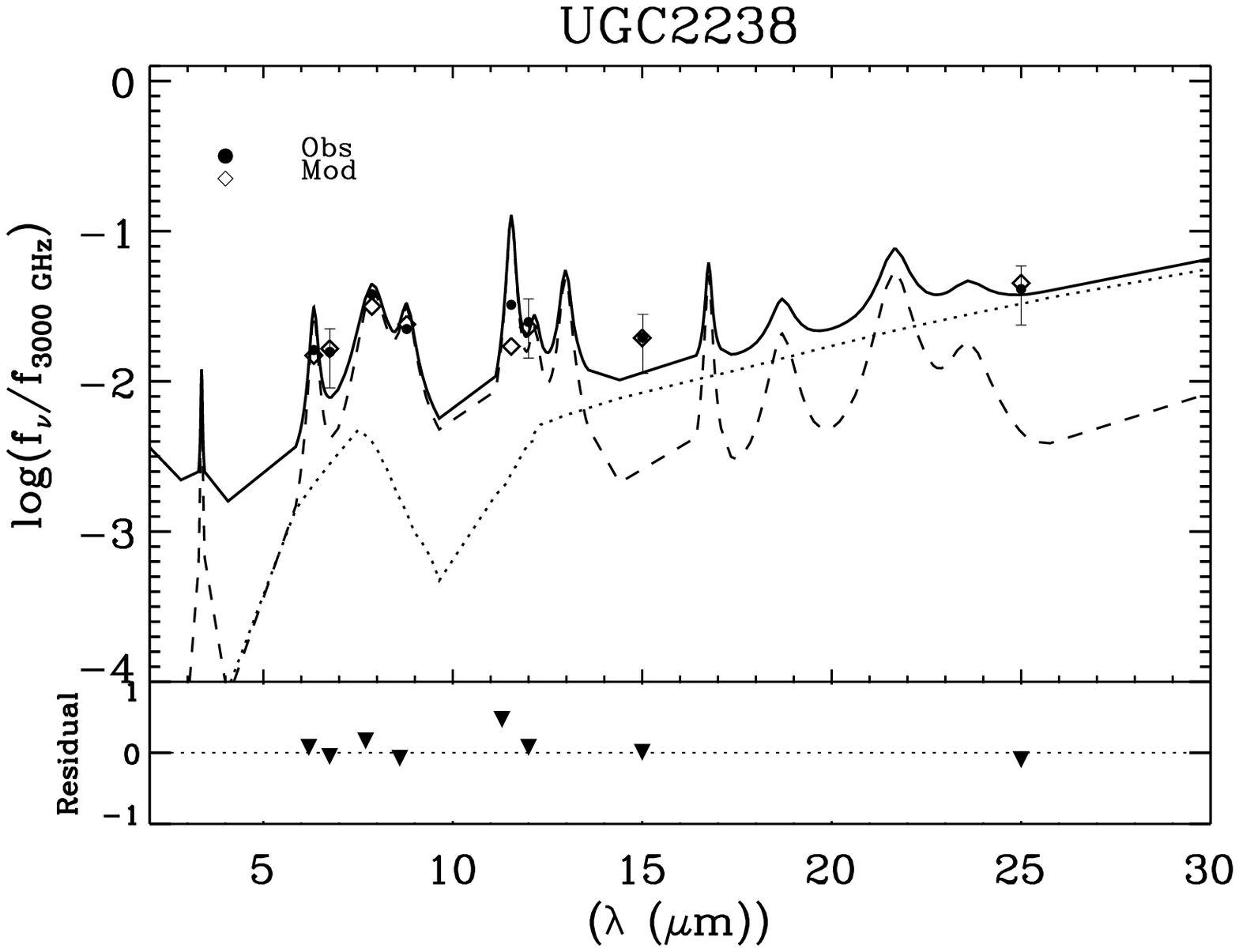}
\includegraphics[angle=0,width=8truecm]{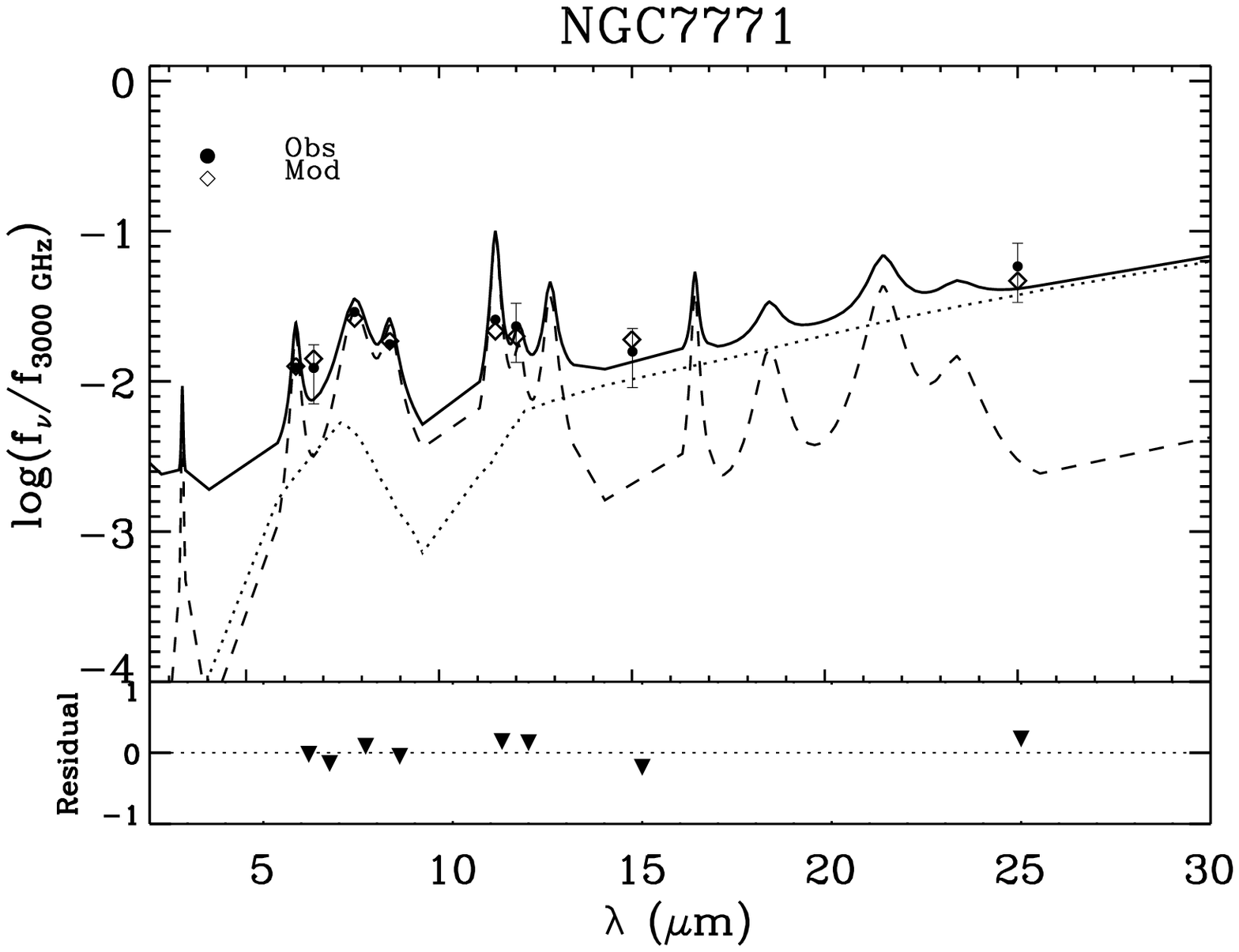}
\includegraphics[angle=0,width=8truecm]{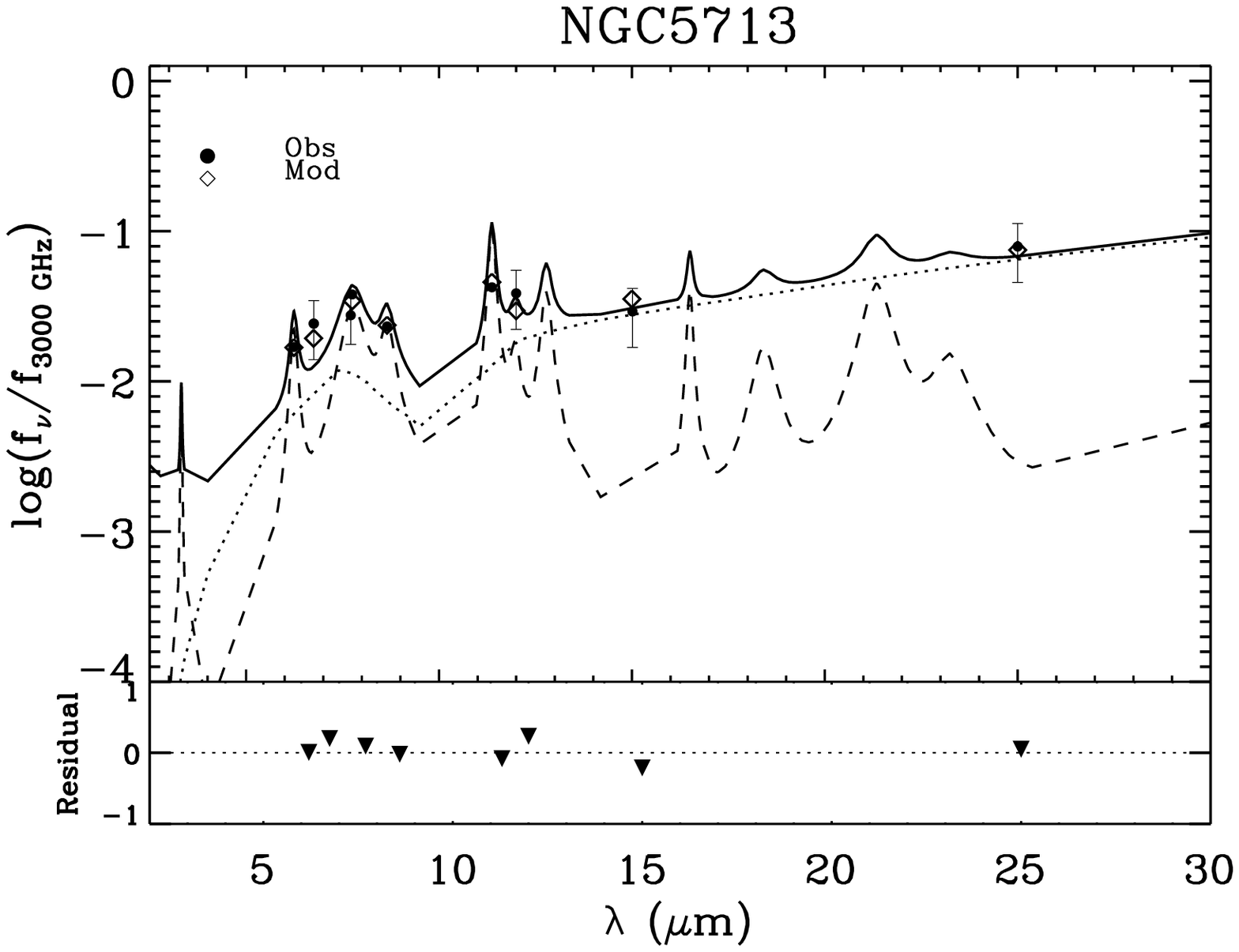}
\caption{Detailed comparison between the observed and model PAH
bands. The solid line is the total emission, the dashed line is
the contribution from the diffuse medium, and the dotted line is
the contribution from the molecular clouds. Filled circles are the
observational data by Lu et al. (2003) and Dale et al. (2000), in
the following order: 6.2$\mu$m,  6.75$\mu$m (ISO LW2), 7.7$\mu$m,
8.6$\mu$m,  11.3$\mu$m, 12$\mu$m (IRAS), 15$\mu$m (ISO LW3) and
25$\mu$m (IRAS). Open rombs are for the model. The residuals
between models and data are given as
$(f_\rmn{data}-f_\rmn{model})/f_\rmn{data}$. In all cases the MIR
continuum is mostly dominated by the molecular clouds emission,
while the bands containing PAHs are initially dominated by the
cirrus component, and then by the molecular contribution, at
increasing $f_\nu (60 \mu \rmn{m}) /f_\nu (100 \mu \rmn{m})$  flux
ratio.}
 \label{fig:pahcalnew}
\end{figure*}

\begin{figure*}
\addtocounter{figure}{-1}
\centering
\includegraphics[angle=0,width=8truecm]{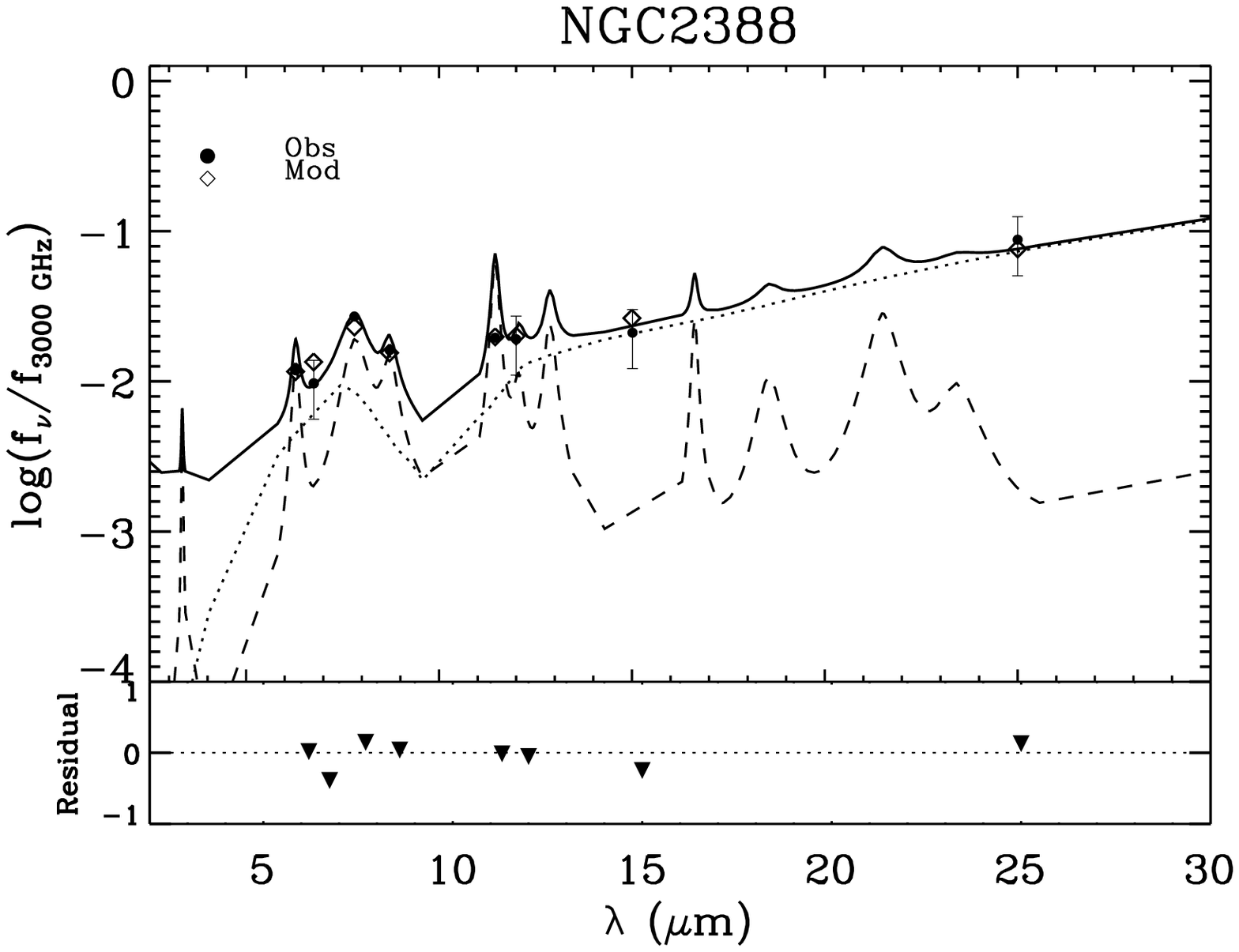}
\includegraphics[angle=0,width=8truecm]{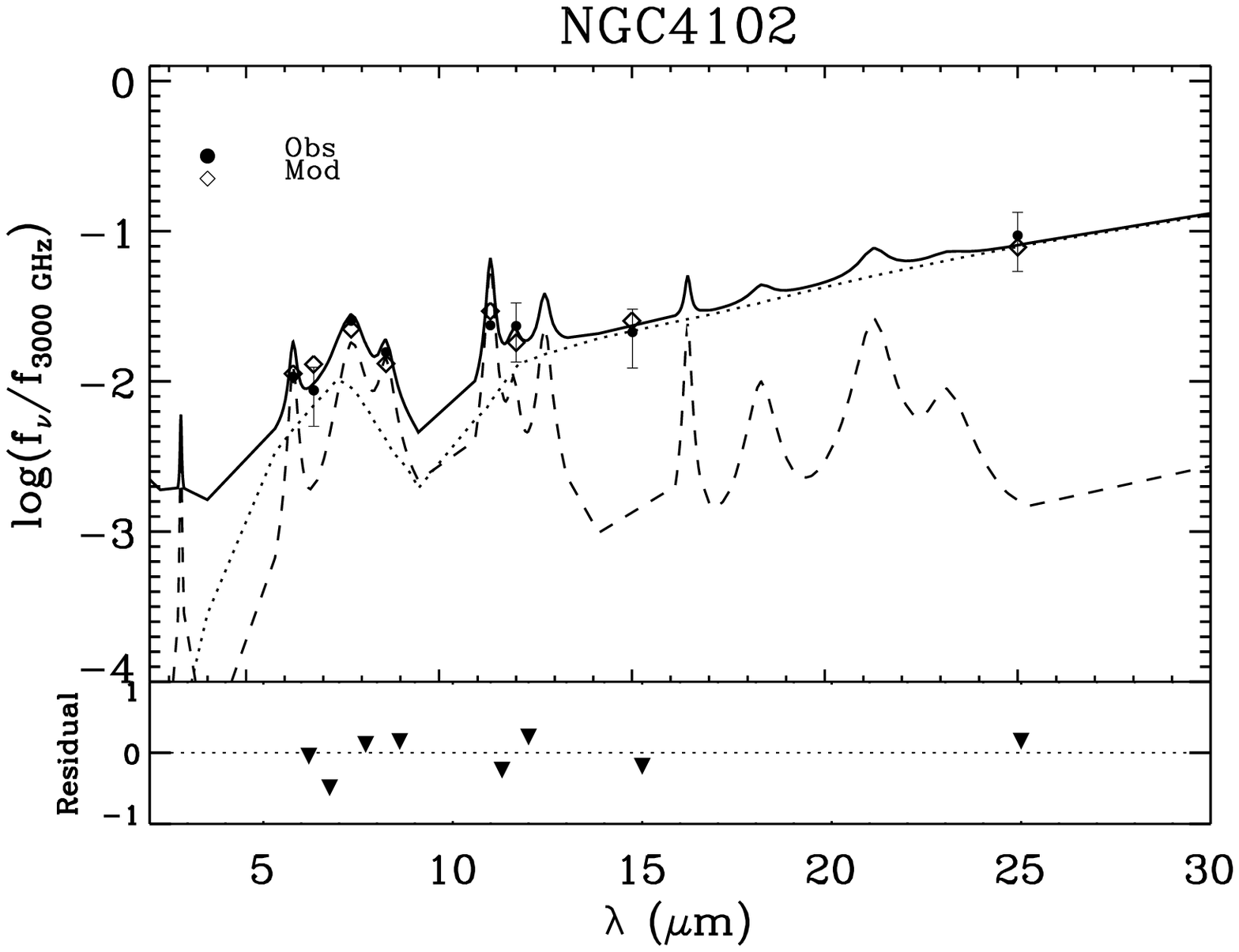}
\includegraphics[angle=0,width=8truecm]{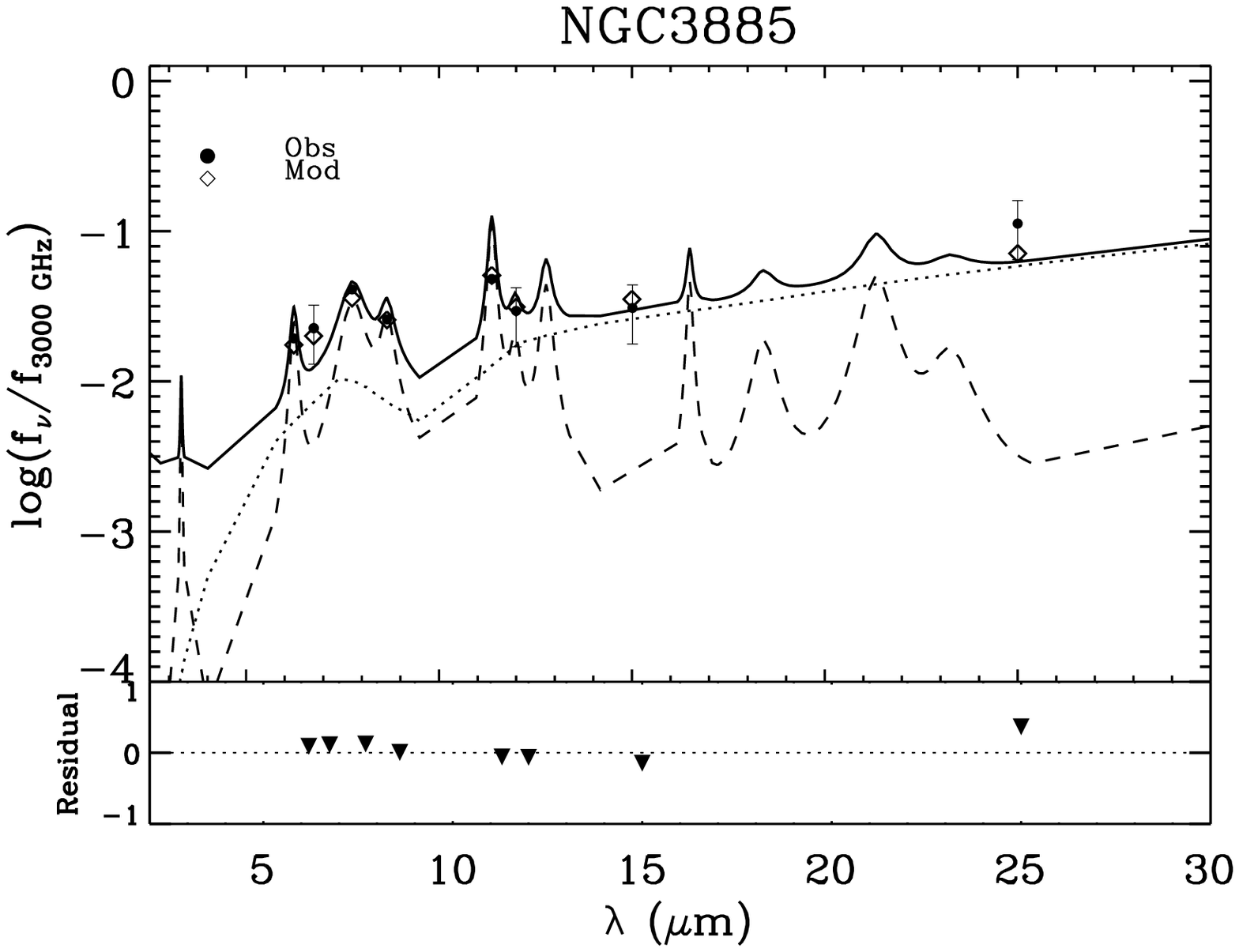}
\includegraphics[angle=0,width=8truecm]{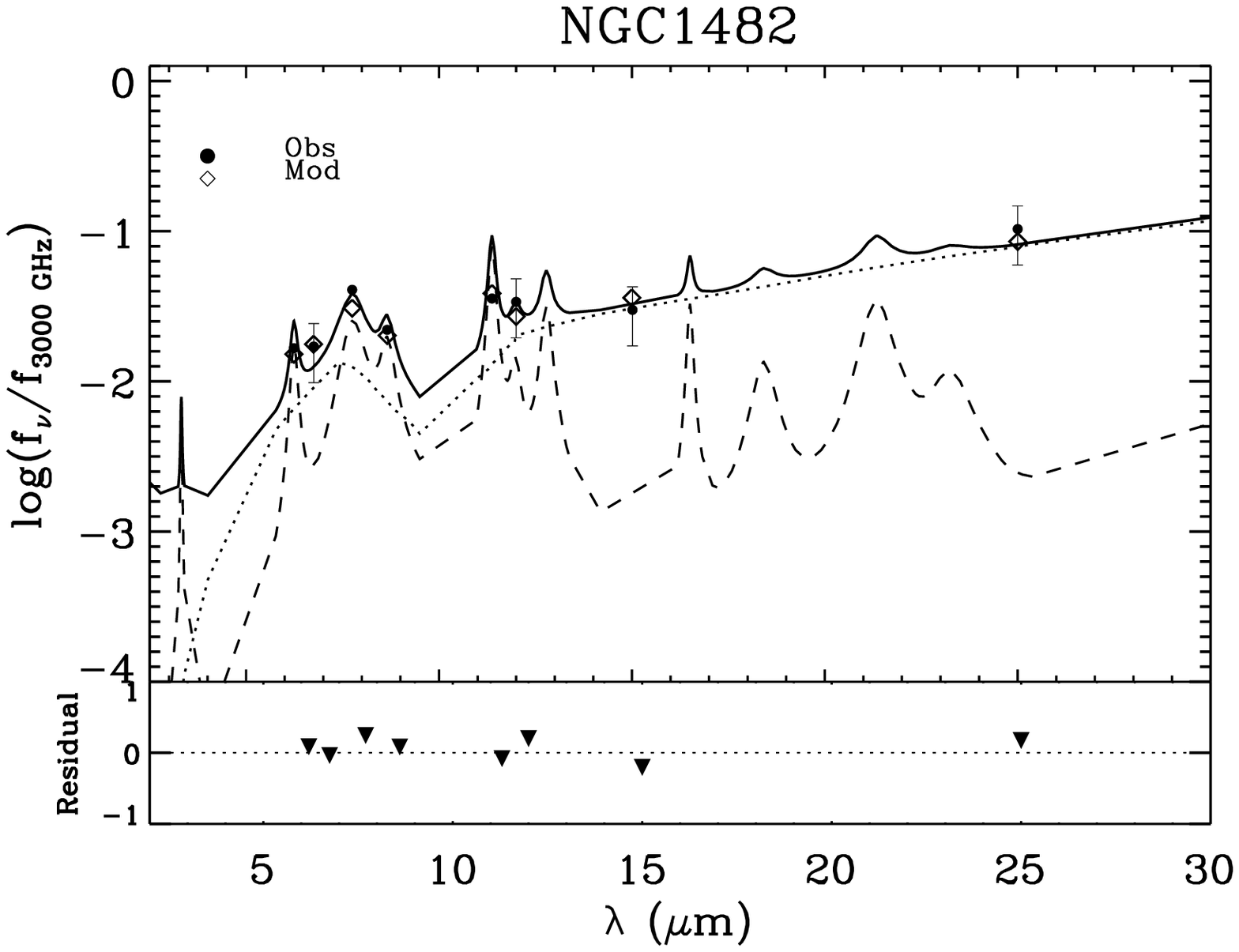}
\includegraphics[angle=0,width=8truecm]{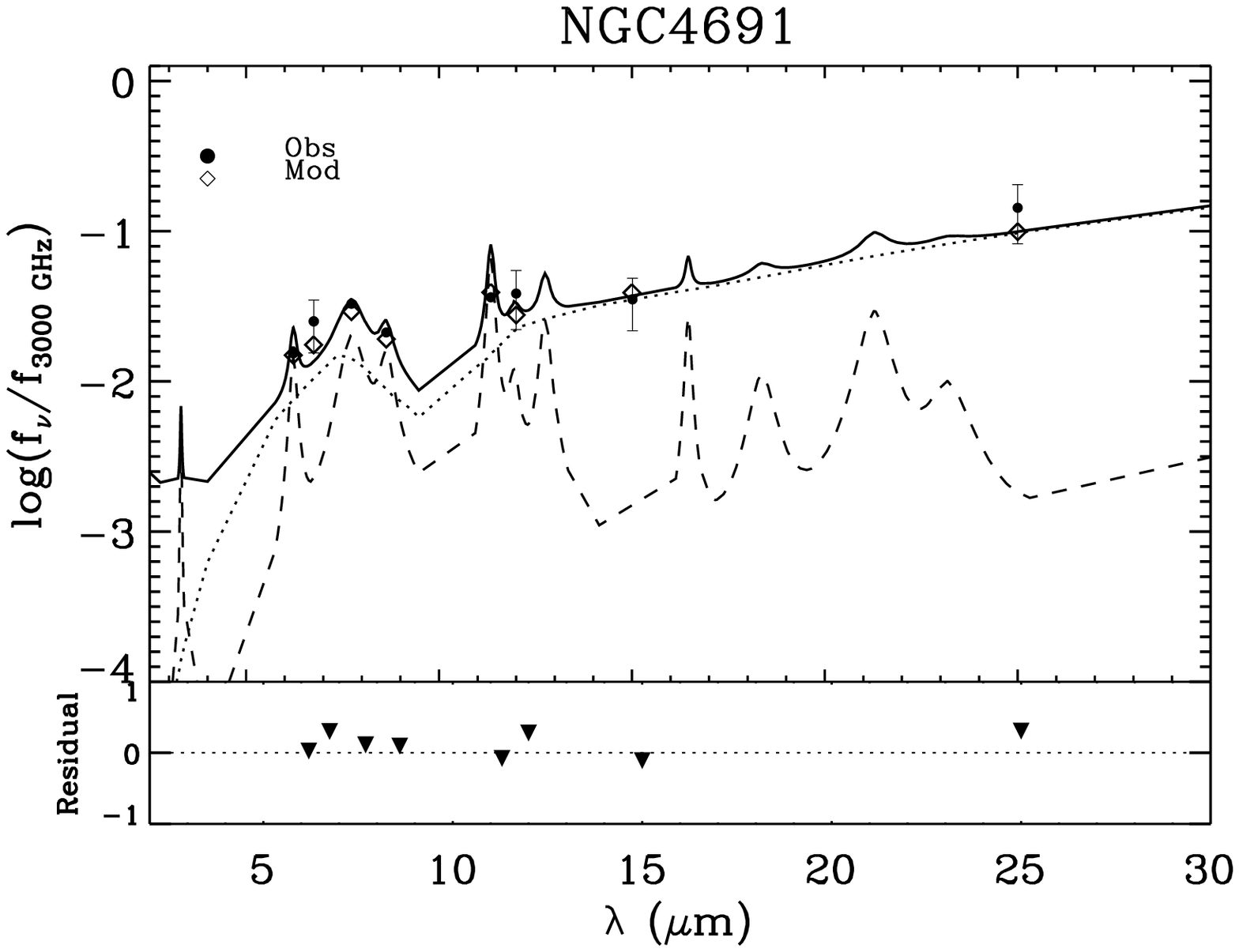}
\includegraphics[angle=0,width=8truecm]{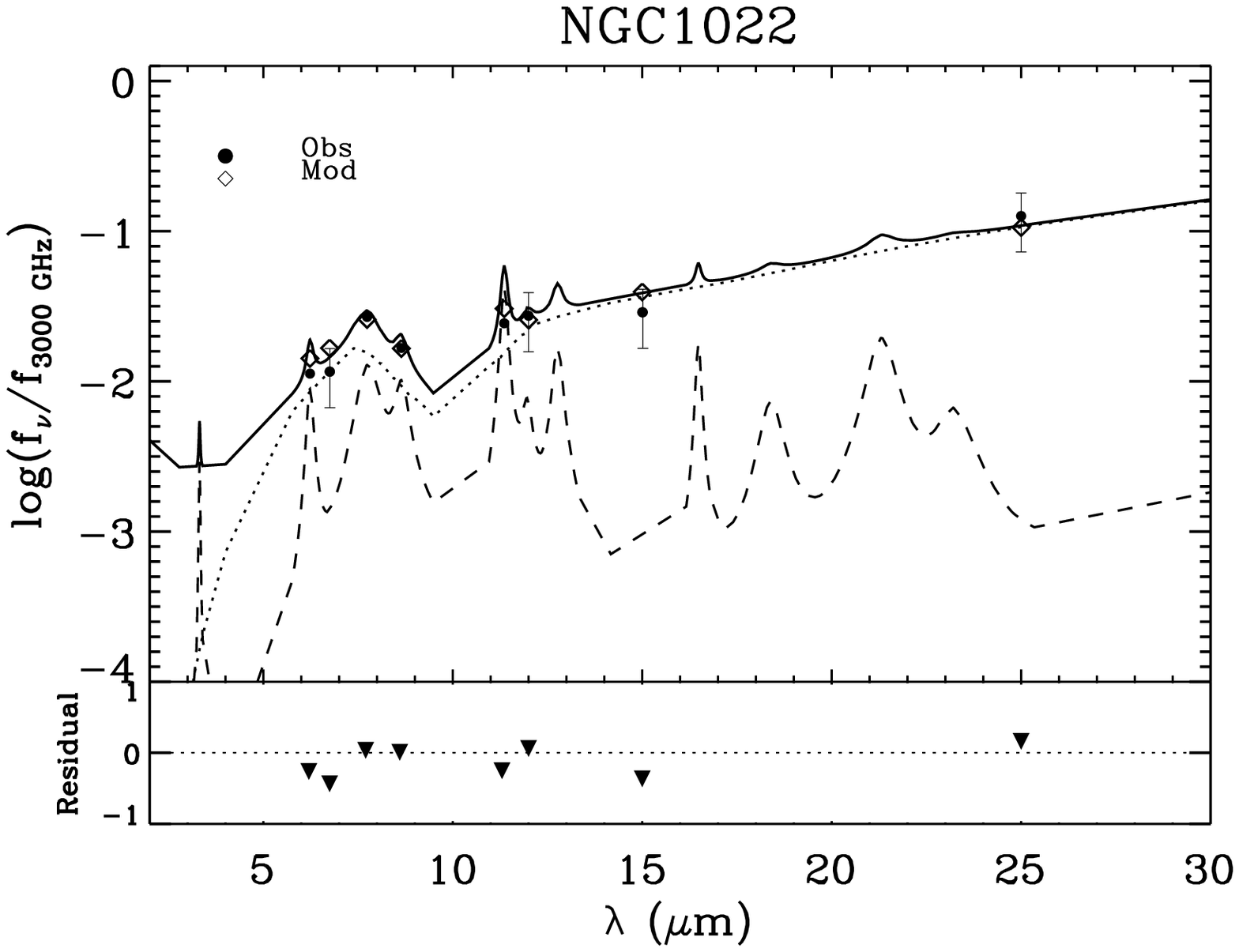}
\includegraphics[angle=0,width=8truecm]{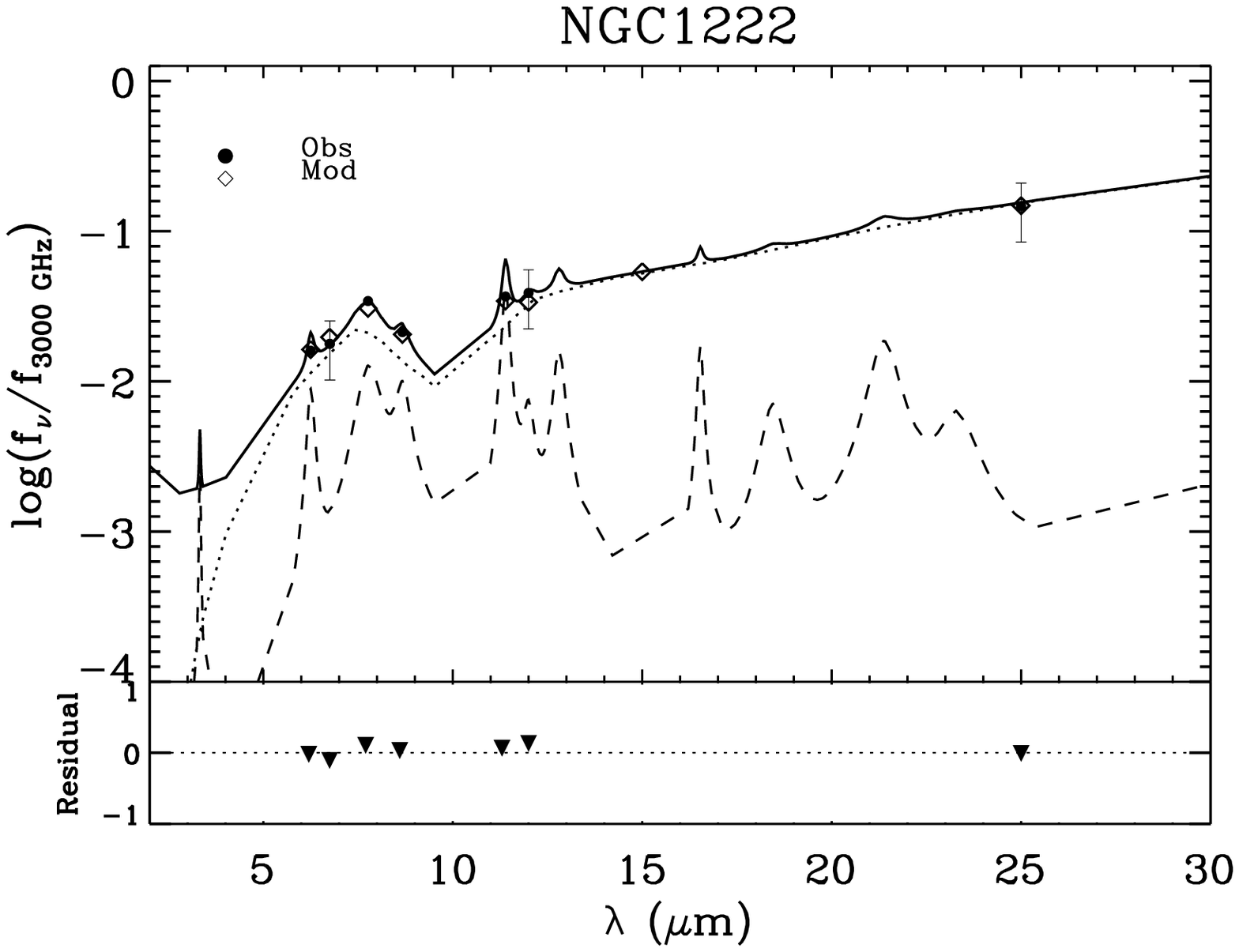}
\includegraphics[angle=0,width=8truecm]{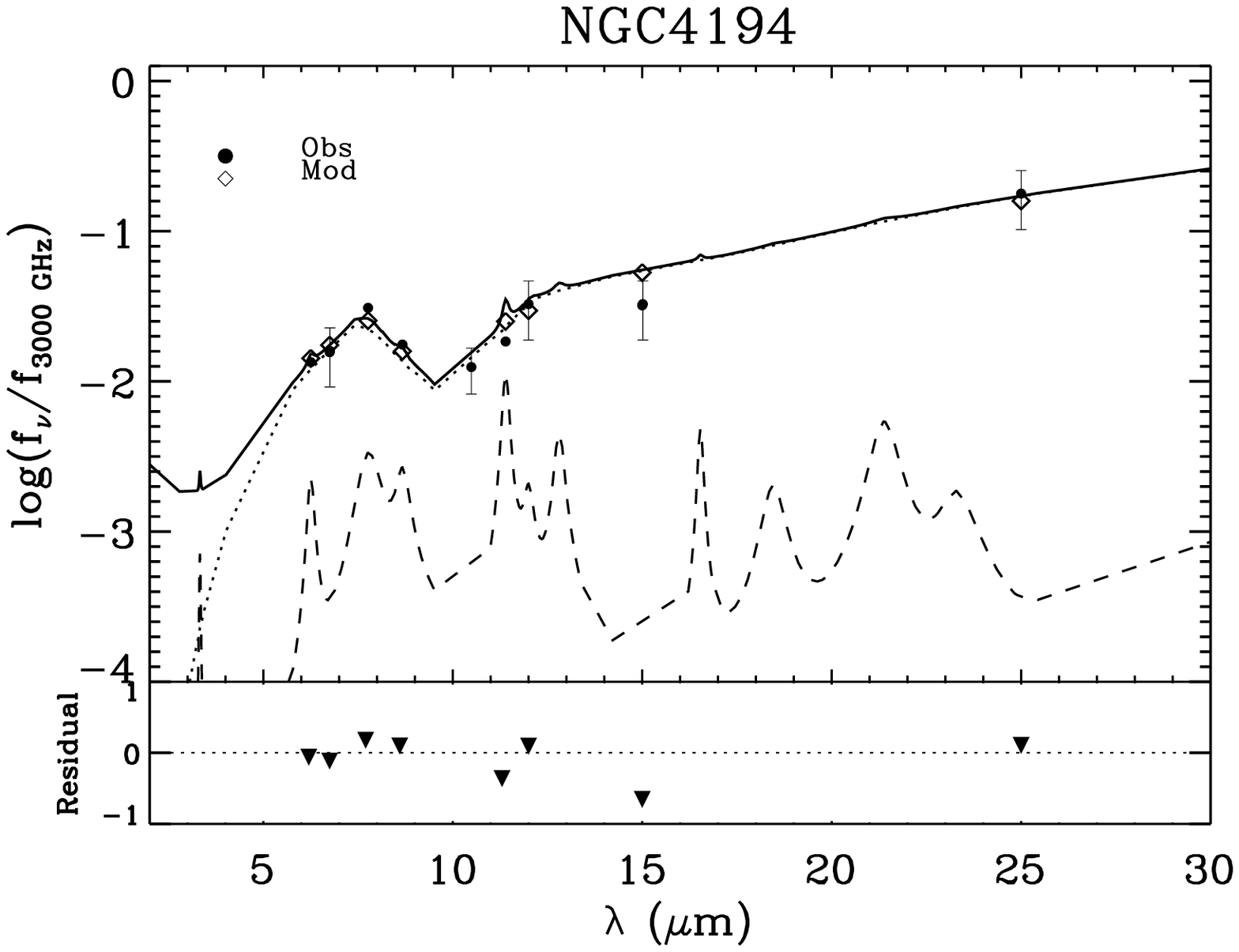}
\caption{Continued}
\end{figure*}

%%%%%%%%%%%%%%%%%%%%%%%%%%%%%%%%%%%%%%%%%%%%%%%%%%%%%%%%%%%%%%%%%%%%%

\section{Comparison with MID Infrared SEDs of selected galaxies}
\label{sec:callu}

In order to test the model with observations we performed a detailed comparison with well observed
MIR SEDs  of a sample of star forming galaxies taken from Lu et al. (2003). This galaxy sample,
observed  with the PHT-S mode of ISOPHOT (5.8 - 11.6~$\mu$m), covers the full range of
morphological types of disk galaxies, from S0 to Im, and is characterized by an increasing ratio of
present day star forming activity to the past time averaged star formation rate (Helou 1986). The
Lu et al. galaxies are  a subset of a larger sample for which ISO (Dale et al. 2000) and IRAS broad
band fluxes are available, and will be discussed in Section~\ref{sec:colcol}.

Among this sample, we selected those galaxies which are classified as HII on the NED\footnote{The
NASA/IPAC Extragalactic Database is operated by the Jet Propulsion Laboratory, California Institute
of Technology, under contract with the National Aeronautics and Space Administration.},
%and {\bf with prominent PAH features easily identified by eye}
and with well defined PAH features, i.e. we considered those objects whose data are incompatible
with an absence of features when including error bars. The selected sub-sample is listed in Table
\ref{tablu}, where galaxies are reported in increasing order $f_\nu (60 \mu \rmn{m}) /f_\nu (100
\mu \rmn{m})$ together with one of the results of the fitting, i.e. if the galaxy is found to be
cirrus dominated (c type in the last column), molecular cloud dominated (m) or of mixed type
respectively. The other properties of these galaxies can be retrieved in Table~\ref{tab:tabdale}.

Lu et al. provide  specific fluxes within narrow square passbands
around the corresponding PAH features, for an aperture of
$24\arcsec \times 24\arcsec$. In order to compare these values
with the global broad-band photometry at 6.75 $\mu$m (LW2) taken
by Dale et al. (2000), we used the aperture coverage factor, $p$,
given by Lu et al. With this factor, the errors in the fit of the
PAH bands by using the 6.75~$\mu$m broadband is always less than
25\%.

The fits of the observed SEDs were performed on a wider wavelength
range (from NIR to radio, using H, J, K bands, LW2, LW3 ISO bands,
12, 25, 60 and 100 $\mu$m IRAS bands, 450, 850 $\mu$m SCUBA bands,
and 8.4, 4.8, and 1.49 GHz bands when the data are available, see
below and Fig. 1) and by taking into account the suitable filter
responses.

We considered galaxy models spanning a wide range of parameters:
star formation history, obscuration times, dust optical
properties, etc. The star formation history was obtained with our
chemical evolution code, adopting a Schmidt-type SFR
($=\nu_\rmn{Sch} M_\rmn{gas}$), with a star formation efficiency
$\nu_\rmn{Sch} = 0.5$ Gyr$^{-1}$, a Salpeter (1955) IMF ($\propto
M^\rmn{-1.35}$) between $0.05-120 \rmn{M_\odot}$. The
observational sample is made of star forming galaxies with a large
range of IR luminosities, revealing the presence of mild to strong
starbursts. Therefore we superimposed to the quiescent component a
starburst phase characterized by an exponential decreasing star
formation rate (with e-folding times $t_\rmn{b}$). The burst is
supposed to start at a galaxy age of 11.95 Gyr, and to involve
10\% of the mass of the gas present at that age. We follow the
evolution of the SEDs at different ages of the superimposed burst
($age_\rmn{b}$), from the early starburst phase to the
post-starburst and quiescent-normal phase.

Among the GRASIL parameters that mostly affect the MIR-FIR range
we recall the optical depth of MCs (we address to it as
$\tau_\rmn{1}$, i.e. at $1 \mu$m), the gas-to-dust mass ratio
($G/D$), the fraction of molecular mass to total gas mass
($M_\rmn{m}/M_\rmn{gas}$), and the escape time-scale of newly born
stars from their parent MCs ($t_\rmn{esc}$). Tables
\ref{tabchevonew} and \ref{tabgranew} list the ranges of
parameters of the model SED library used to perform the fits.

\begin{table}
\centering \caption{Parameters for the star formation history of
the model library. M$_\rmn{low}$ and M$_\rmn{up}$ are the adopted
mass limits of the Salpeter IMF; $\nu_\rmn{Sch}$ is the efficiency
of the Schmidt-type SFR; $t_\rmn{b}$ indicates the e-folding time
of the exponential burst superimposed to the quiescent star
formation; age$_\rmn{b}$ indicates the ages, from the beginning of
the burst, at which the SEDs are computed; \%$M_\rmn{b}$ is the
percentage of the mass of gas of the galaxy models involved in the
burst at an age of 11.95 Gyr. See text for more details.}
\label{tabchevonew}
\begin{tabular}{cccccc} \hline
M$_\rmn{low}$&M$_\rmn{up}$&$\nu_\rmn{Sch}$  &  $t_\rmn{b}$   &  $\log$(age$_\rmn{b}$)  &  \% $M_\rmn{b}$\\
$\rmn{M_\odot}$&$\rmn{M_\odot}$&(Gyr$^{-1}$) &  (Myr)   &  (yr)               &  \\
\hline
0.05&120&0.5          &  10 - 50 &  6.0 - 8.2          & 10 \\
\hline
\end{tabular}
\end{table}

\begin{table}
\centering
 \caption{GRASIL parameters for the SED model library. $G/D$ is the gas to dust mass ratio;
 $t_\rmn{esc}$ is the escape time scale of newly born stars from their parent MCs; $\frac{M_\rmn{m}}{M_\rmn{tot}}$
 is the fraction of gas in MCs; $\tau_{1}$ is the $1 \mu$m optical depth of MCs}.
  \label{tabgranew}

\begin{tabular}{cccc} \hline

 $G/D$ & $t_\rmn{esc}$ & $\frac{M_\rmn{m}}{M_\rmn{tot}}$ & $\tau_{1}$ \\
     &   (Myr)     &
    &\\
  \hline
 100   & 10 - 50   & 0.1 - 0.9                           &   4 - 180 \\

\hline
\end{tabular}
\end{table}

In Fig. \ref{fig:totalsed} we show some examples of the full SED
fittings we performed to the galaxy sample of Table \ref{tablu}.
The galaxies shown in the figure are representative of the range
of $f_\nu (60 \mu \rmn{m}) /f_\nu (100 \mu \rmn{m})$ values
covered by the sample.

In Fig. \ref{fig:pahcalnew} we show the detailed comparison in the
MIR range between out best fitting models and the data for all the
galaxies reported in Table \ref{tablu}. The points in the figure
refer to the specific fluxes at  6.2$\mu$m, 6.75$\mu$m (ISO LW2),
7.7$\mu$m, 8.6$\mu$m, 11.3$\mu$m, 12$\mu$m (IRAS), 15$\mu$m (ISO
LW3) and 25$\mu$m (IRAS). In all cases the predicted PAH features
compare very well with the observations being generally within the
quoted standard deviations. We also notice, from Table \ref{tablu}
that, at increasing $f_\nu (60 \mu \rmn{m}) /f_\nu (100 \mu
\rmn{m})$ ratio, the SEDs shift from cirrus dominated to MC
dominated, without any appreciable trend in the goodness of the
fit. Given the wide range of the luminosity ratios
$L_\rmn{FIR}/L_\rmn{B}$ spanned by the sample galaxies, we
conclude that the model is robust and able to reproduce the
continuum SEDs as well as the MIR spectral features, over a wide
range of relative star formation activity, from mild to strong
starbursts.

\section{Comparison with ISO-IRAS color-color plots}
\label{sec:colcol}

We now compare our models with a larger data set of star-forming galaxies for which only broad-band
ISO and IRAS fluxes are available.

\subsection{The data sample}

This observational sample is taken from Dale et al. (2000). They
published global ISO fluxes for a large sample of galaxies, for
which also IRAS data are available. These galaxies were observed
within the ISO Key Project on the Interstellar Medium of Normal
Galaxies (Helou et al. 1996). The parent sample for this project
has the original criterion of $f_\nu (60 \mu$m)$\gsim$ 3 Jy, a
published redshift, and no active galactic nucleus or Seyfert
classification in the NED, although two galaxies of the sample,
NGC4418 and MRK331, show some evidences of hosting an AGN. They
span the full range of morphologies, mostly spirals and irregular
galaxies, but including two elliptical galaxies; FIR luminosities
are $\sim 10^8 $- $10^{12}$ L$_\odot$; FIR colors are $f_\nu (60
\mu \rmn{m}) /f_\nu (100 \mu \rmn{m}) \sim 0.25 - 1.6 $, and
infrared to blue luminosity ratios $L_\rmn{FIR}/L_\rmn{B} \sim
0.05 - 50$. They observed 61 galaxies with the broadband filters,
LW2 (6.75 $\mu$m, $\delta \lambda = 3.5$ $\mu$m) and LW3 (15.0
$\mu$m, $\delta \lambda =6.0$~$\mu$m) of ISOCAM (Cesarsky et al.
1996). For this work, we did not include 7 galaxies (NGC1222,
UGC2855, NGC2366, NGC4490, NGC5866, NGC6822, and NGC6946), that
either have only lower limits in their MIR observations, or are
not detected in any LW filter.

For the majority of the galaxies we have also collected, from
 literature, total fluxes on radio emission at 6.3 and
21 cm (Condon et al. 1990, 1991, 1995), that will be used for the
characterization of the starburst evolution. All these data are
shown in Table \ref{tab:tabdale}.

\begin{figure*}
\centering
\includegraphics[angle=0,width=12truecm]{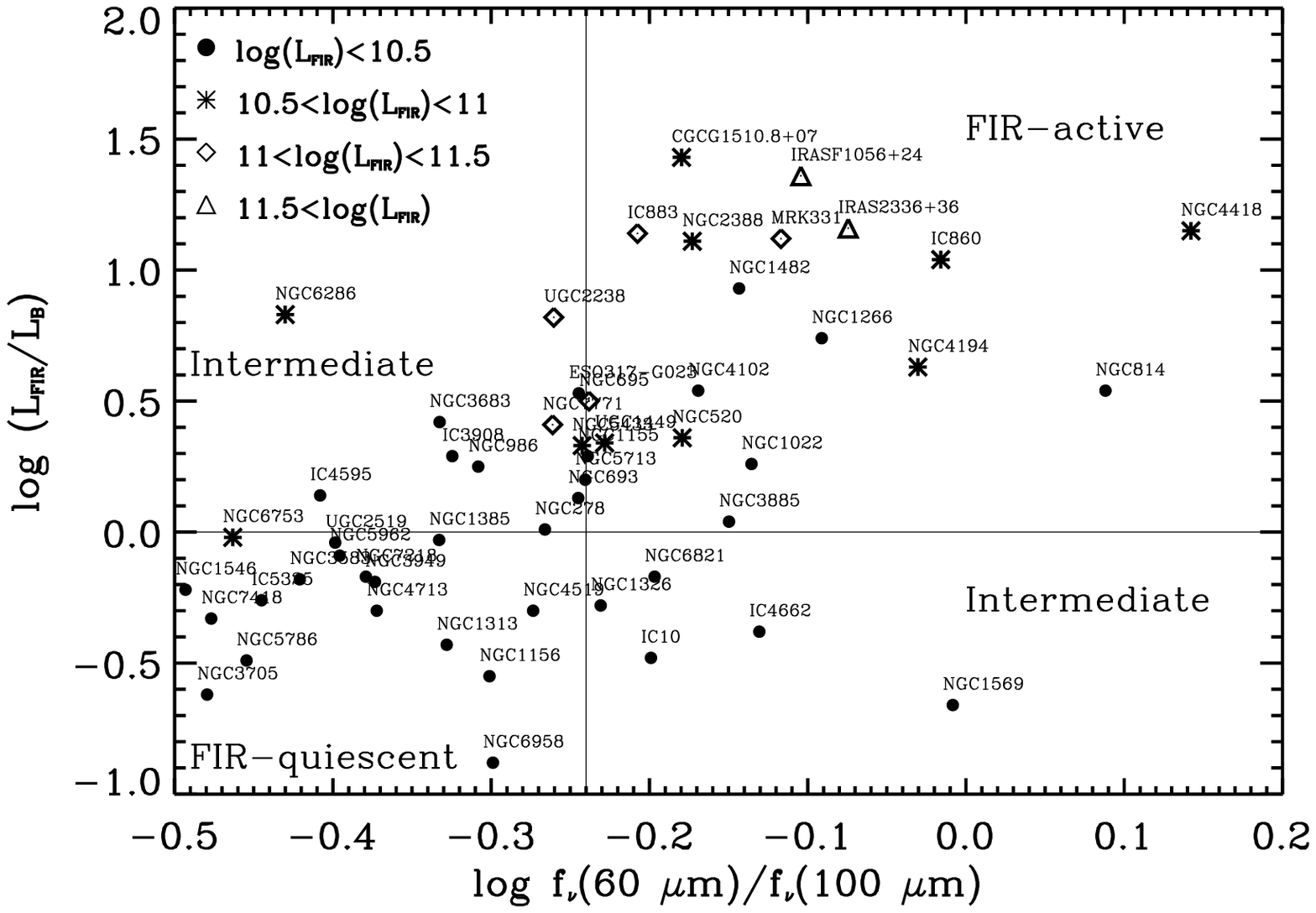}
\caption{Log($L_\rmn{FIR}/L_\rmn{B}$) versus log $f_\nu (60 \mu
\rmn{m}) /f_\nu (100 \mu \rmn{m})$ for the galaxies of the Dale et
al. (2000) sample. Different symbols indicate different FIR
luminosities.} \label{fig:dalesample}
\end{figure*}

Fig. \ref{fig:dalesample} shows the location of the galaxies in the regions corresponding to
different star formation activities, following the definition by Lu et al. (2003). About 43\%
galaxies of the sample have $\log(L_\rmn{FIR}/L_\rmn{B}) \geq$0 and $\log f_\nu (60 \mu \rmn{m})
/f_\nu (100 \mu \rmn{m}) \geq-0.24$, and are considered to be strong starbursts. About 31\%
galaxies of the sample fall in the opposite quadrant and are considered to be FIR-quiescent
galaxies, while the remaining 26\%, populating the other two quadrants, are considered to be in a
mild starburst phase. In summary the sample of Dale et al. is well suited to study a mild to strong
starburst environment.

\begin{table*}
\centering \scriptsize{ \caption{Properties of the Dale et al.
(2000) sample we use for our analysis. Radio data are from Condon
et al. (1990, 1991, 1995).} \label{tab:tabdale}
\begin{tabular}{llrrrrrrrrrr} \hline

   &NAME &$12\mu$m &$25\mu$m &$60\mu$m&$100\mu$m&$6.75\mu$m&$15\mu$m& $D_\rmn{L}$ &$log(\frac{L_\rmn{FIR}}{L_\rmn{B}})$&  6.3 cm
   &21 cm\\
   &&IRAS&IRAS&IRAS&IRAS&ISO-LW2&ISO-LW3\\
* &  &Jy & Jy & Jy &Jy&Jy &Jy &Mpc&&Jy&Jy\\
\hline
1&IC10 &4.88& 13.95& 112.92 &178.56& 2.18&3.06 &0.7 &-0.48& 0.202& 0.298 \\
2 &NGC278& 1.63& 2.57& 25.05&46.22& 1.12& 1.18 &11.8& 0.01& 0.063&0.138\\
3 &NGC520& 0.76& 2.84 &31.10&46.99&0.52&0.86&31.6 &0.36& 0.126& 0.177\\
4&NGC693&0.29&0.55&6.73&11.83&0.18&0.18&22.1&0.13&0.033&0.053\\
5&NGC695&0.48& 0.86& 7.87& 13.62&0.25& 0.36& 131.8& 0.50& 0.025&  0.076\\
6 &UGC1449& 0.31& 0.56& 4.96&8.39& 0.18&0.17 &75.5& 0.34&  -&-\\
7 &NGC814&0.19&1.01&4.41&3.60& 0.05& 0.22& 21.4&0.54&-&0.007\\
8 &NGC986&1.41&3.65&25.14&51.11&0.51&1.07&25.2&0.25&0.040&0.110\\
9 &NGC1022&0.75&3.29&19.83&27.09&0.31&0.77&19.4&0.26&0.018&0.045\\
10&UGC2238&0.34&0.53&8.40&15.30&0.24&0.30&86.8&0.82&0.040&0.217 \\
11&NGC1155&0.21&0.47&2.89&5.01&0.05&0.09&60.3&0.29&0.018&-\\
12&NGC1156&0.17&0.55&5.24&10.48&0.09&0.14&6.4&-0.55&-&0.022 \\
13&NGC1222&0.51&2.29&13.07&15.45&0.27&-&32.6&0.44&0.029&0.055\\
14&UGC2519&0.25&0.34&2.98&7.46&0.12&0.17&34.6&-0.04&-&-\\
15&NGC1266&0.14&1.23&13.32&16.43&0.04&0.15&29.1&0.74&-&0.110\\
16&NGC1313&1.70&3.75&45.69&97.25&0.69&0.84&3.7&-0.43&0.123&-\\
17&NGC1326&0.38&0.86&8.17&13.90&0.18&0.32&16.2&-0.28&0.035&0.031\\
18&NGC1385&1.19&2.02&17.30&37.23&0.72&0.76&18.4&-0.03&0.078&0.162 \\
19&UGC2855&2.93&4.86&42.39&90.49&-&-&18.7&0.31&0.135&-\\
20&NGC1482&1.54&4.67&33.45&46.53&0.78&1.37&24.0&0.93&0.125&0.216\\
21&NGC1546&0.62&0.79&7.21&22.44&0.42&0.56&14.1&-0.22&-&- \\
22&NGC1569&1.23&8.98&54.25&55.30&0.32&1.43&2.5&-0.66&0.202&0.411\\
23&NGC2388&0.51&2.07&17.01&25.33&0.23&0.50&54.8&1.11&0.027&0.073\\
24&NGC2366&0.21&1.05&4.85&5.04&0.02&0.17&2.9&-0.83&-&0.025\\
25&ESO317-G023&0.34&0.88&13.50&23.71&0.18&0.27&34.7&0.53&-&0.075 \\
26&IRASF1056+24&0.17&1.19&12.08&15.36&0.09&0.27&171.4&1.36&0.020&0.050\\
27&NGC3583&0.63&0.78&7.08&18.66&0.29&0.41&29.2&-0.18&0.030&0.053 \\
28&NGC3620&1.29&4.71&46.80&67.26&0.63&1.06&19.0&5.00&0.120&-\\
29&NGC3683&1.06&1.53&13.61&29.27&0.64&0.76&24.2&0.42&0.045&0.099\\
30&NGC3705&0.38&0.44&3.72&11.22&0.19&0.27&11.9&-0.62&-&0.021 \\
31&NGC3885&0.46&1.41&11.66&16.46&0.33&0.45&20.8&0.04&-&0.046\\
32&NGC3949&0.82&1.37&11.28&26.65&0.42&0.50&11.6&-0.19&0.037&0.100 \\
33&NGC4102&1.72&7.05&48.10&71.01&0.66&1.61&12.4&0.54&0.070&0.261 \\
34&NGC4194&0.83&4.53&23.72&25.44&0.41&0.84&34.8&0.63&0.039&0.102\\
35&NGC4418&1.00&9.69&43.89&31.64&0.24&1.56&27.3&1.15&-&0.039\\
36&NGC4490&1.86&4.20&45.90&76.93&-&-&8.2&-0.20&0.295&0.779\\
37&NGC4519&0.36&0.55&3.74&7.02&0.11&0.23&15.1&-0.30&0.006&0.009\\
38&NGC4713&0.24&0.17&4.60&10.84&0.15&0.21&7.4&-0.30&-&0.038\\
39&IC3908&0.44&0.87&8.09&17.08&0.44&0.43&15.4&0.29&0.440&0.079\\
40&IC860&0.10&1.31&17.93&18.60&0.02&0.06&44.7&1.04&4.000&0.033\\
41&IC883&0.25&1.36&15.44&24.90&0.16&0.25&94.0&1.14&0.046&0.101\\
42&NGC5433&0.27&0.70&6.62&11.57&0.18&0.30&59.0&0.33&-&0.059\\
43&NGC5713&1.30&2.84&21.89&38.09&0.88&1.06&24.7&0.20&0.081&0.154\\
44&NGC5786&0.36&0.76&5.26&14.98&0.30&0.39&37.9&-0.49&-&0.040\\
45&NGC5866&0.36&0.34&5.21&17.11&-&-&11.4&-0.61&-&0.021\\
46&CGCG1510.8+07&0.05&0.83&20.84&31.52&0.04&0.08&52.4&1.43&-&-\\
47&NGC5962&0.74&1.03&8.89&22.11&0.37&0.51&27.3&-0.09&0.036&0.077\\
48&IC4595&0.71&0.73&7.05&18.04&0.30&0.39&42.9&0.14&0.035&0.062\\
49&NGC6286&0.42&0.56&8.22&22.13&0.17&0.20&76.5&0.83&0.050&0.156\\
50&IC4662&0.30&1.27&8.81&11.90&0.05&0.17&2.1&-0.38&0.029&-\\
51&NGC6753&0.60&0.72&9.43&27.40&0.38&0.51&40.4&-0.02&0.035&- \\
52&NGC6821&0.14&0.31&3.63&5.71&0.06&0.13&22.5&-0.17&-&-\\
53&NGC6822&0.84&6.63&58.86&130.32&0.10&0.31&0.7&-0.58&-&0.048\\
54&NGC6946&15.17&23.34&167.72&362.66&11.19&11.49&5.5&-0.34&0.493&1.395 \\
55&NGC6958&0.16&0.20&1.00&1.99&0.03&0.03&36.4&-0.88&-&0.018\\
56&NGC7218&0.28&0.56&4.67&11.18&0.21&0.28&23.8&-0.17&-&-\\
57&NGC7418&0.63&0.69&5.38&16.13&0.23&0.29&19.4&-0.33&-&0.046\\
58&IC5325&0.48&0.70&5.15&14.35&0.24&0.31&19.7&-0.26&-&-\\
59&IRAS2336+36&0.13&0.88&7.44&8.83&0.02&0.11&261.2&1.16&-&-\\
60&NGC7771&0.87&2.18&20.50&37.40&0.46&0.59&60.0&0.41&0.060&0.124 \\
61&MRK331&0.55&2.39&18.04&23.61&0.21&0.52&76.8&1.12&0.028&0.068\\
\hline  \hline
\end{tabular}}
\end{table*}
\normalsize

\begin{figure*}
\centering
\includegraphics[angle=0,width=8.8truecm]{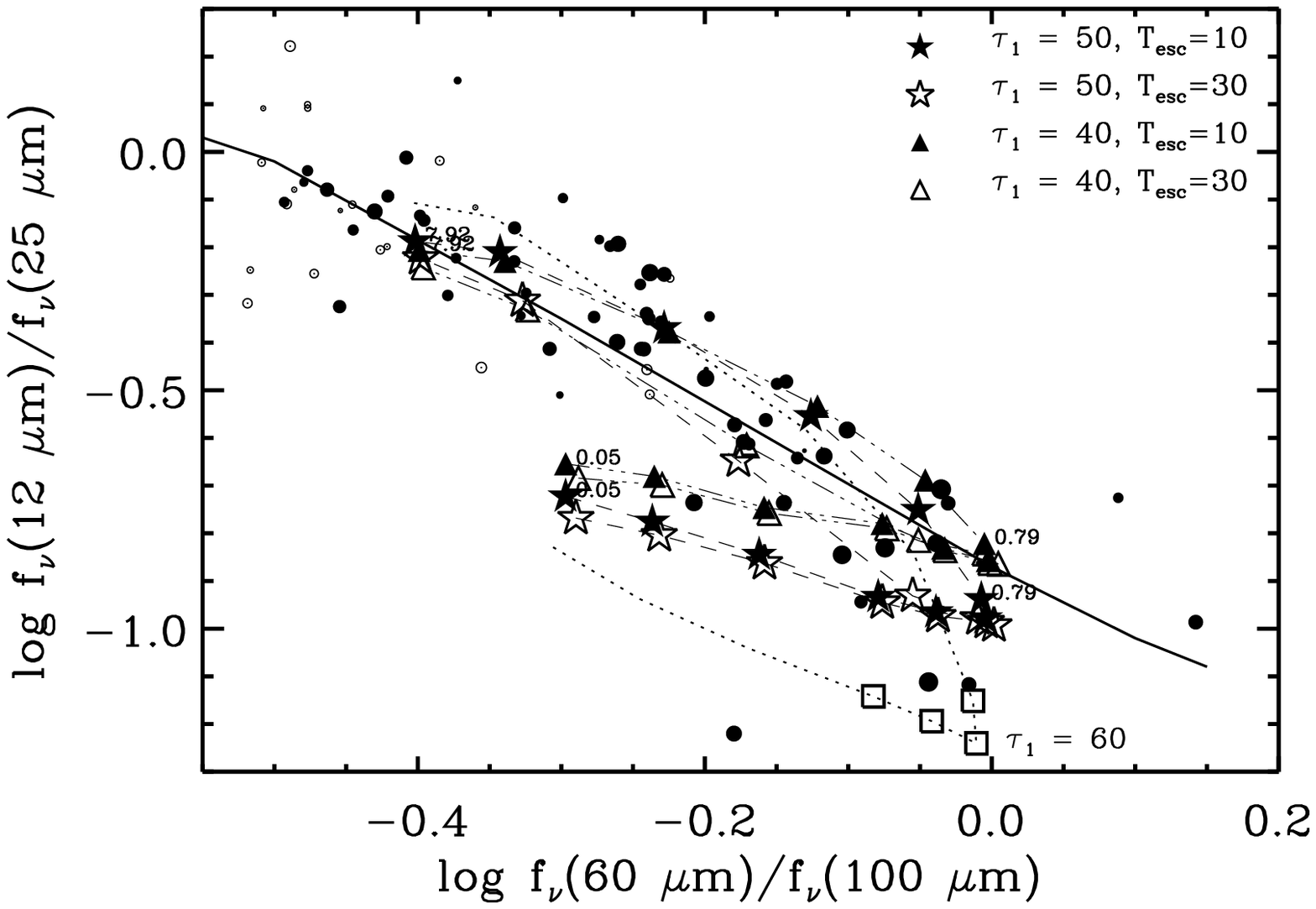}
\includegraphics[angle=0,width=8.8truecm]{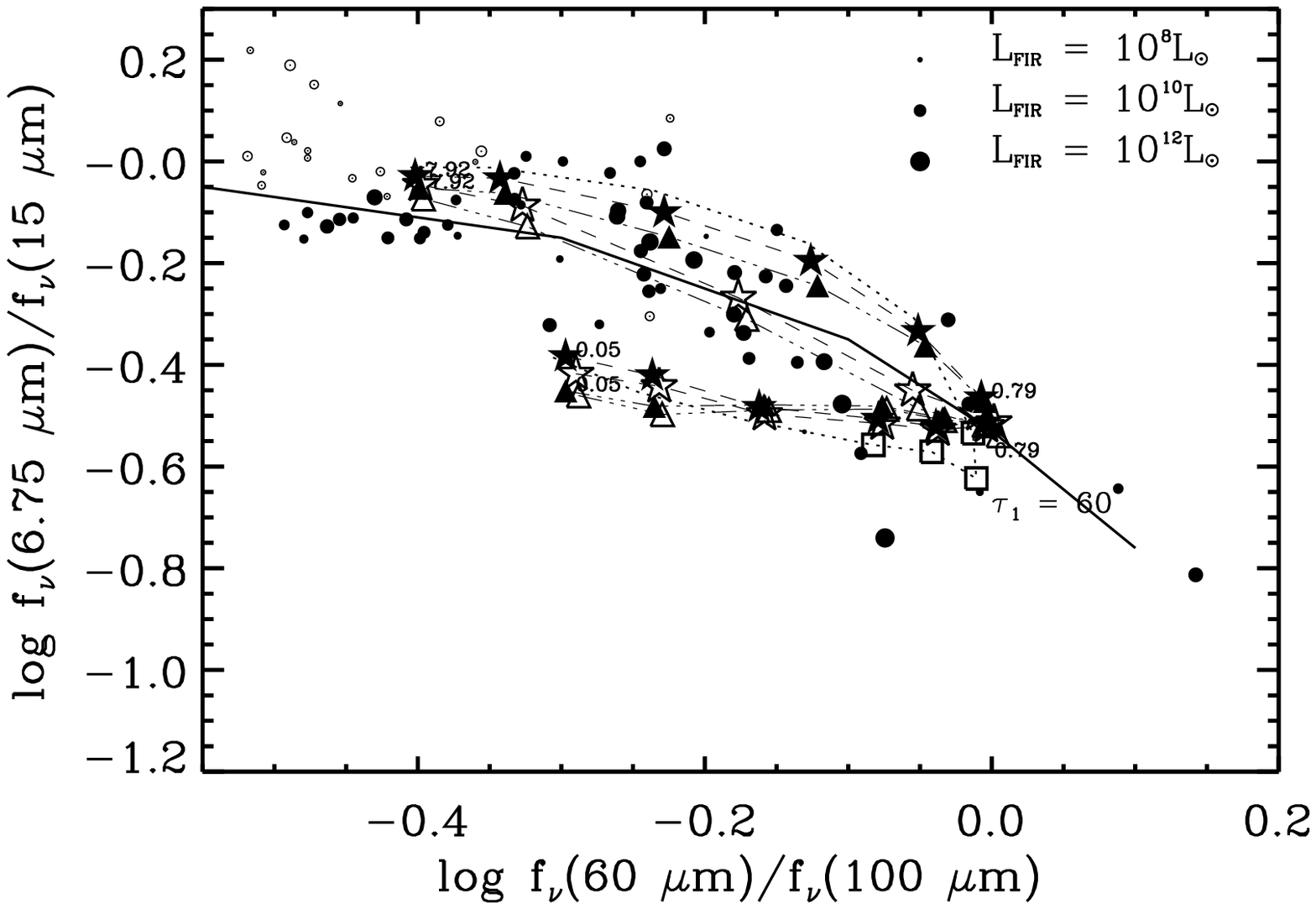}
\includegraphics[angle=0,width=8.8truecm]{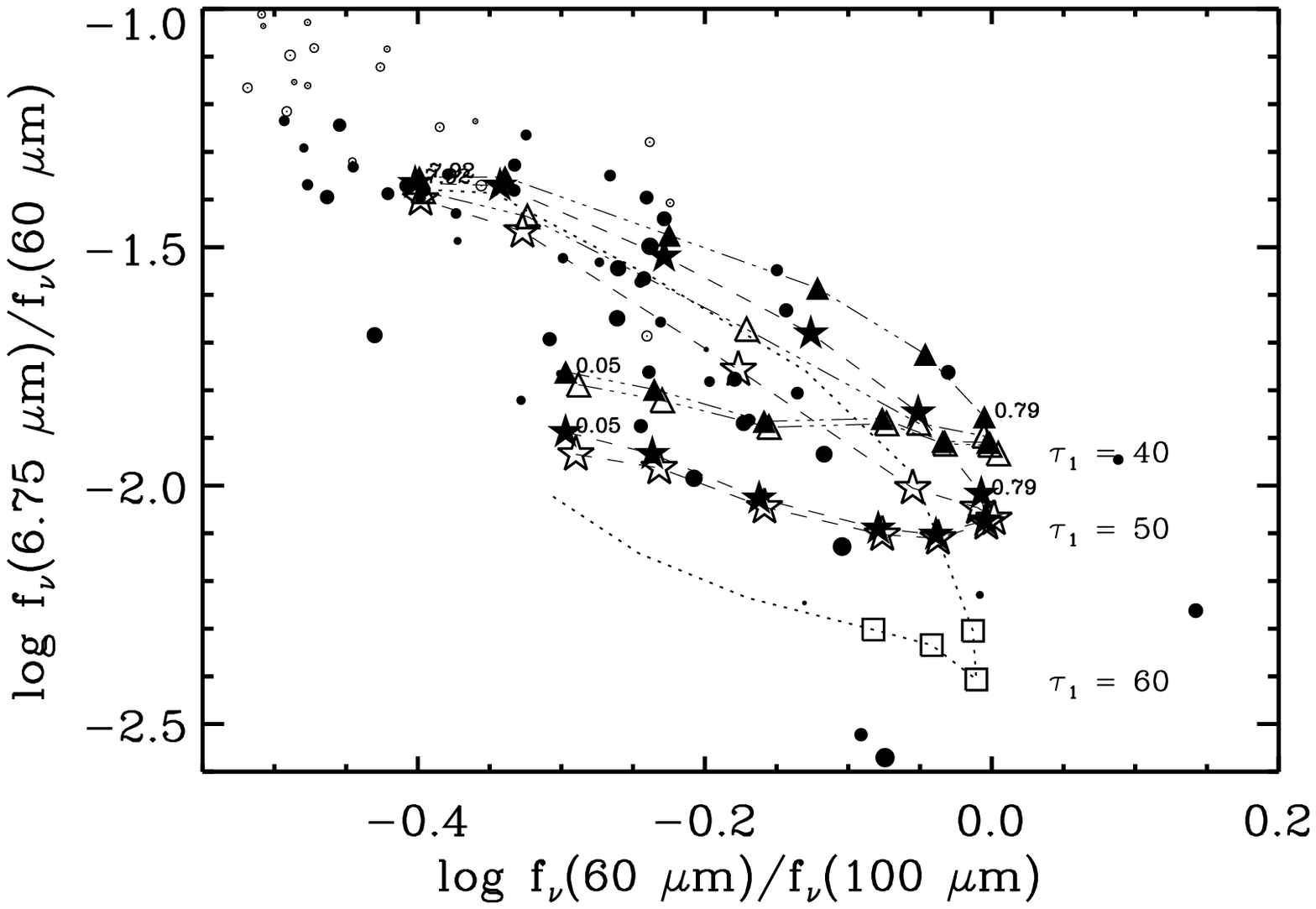}
\includegraphics[angle=0,width=8.8truecm]{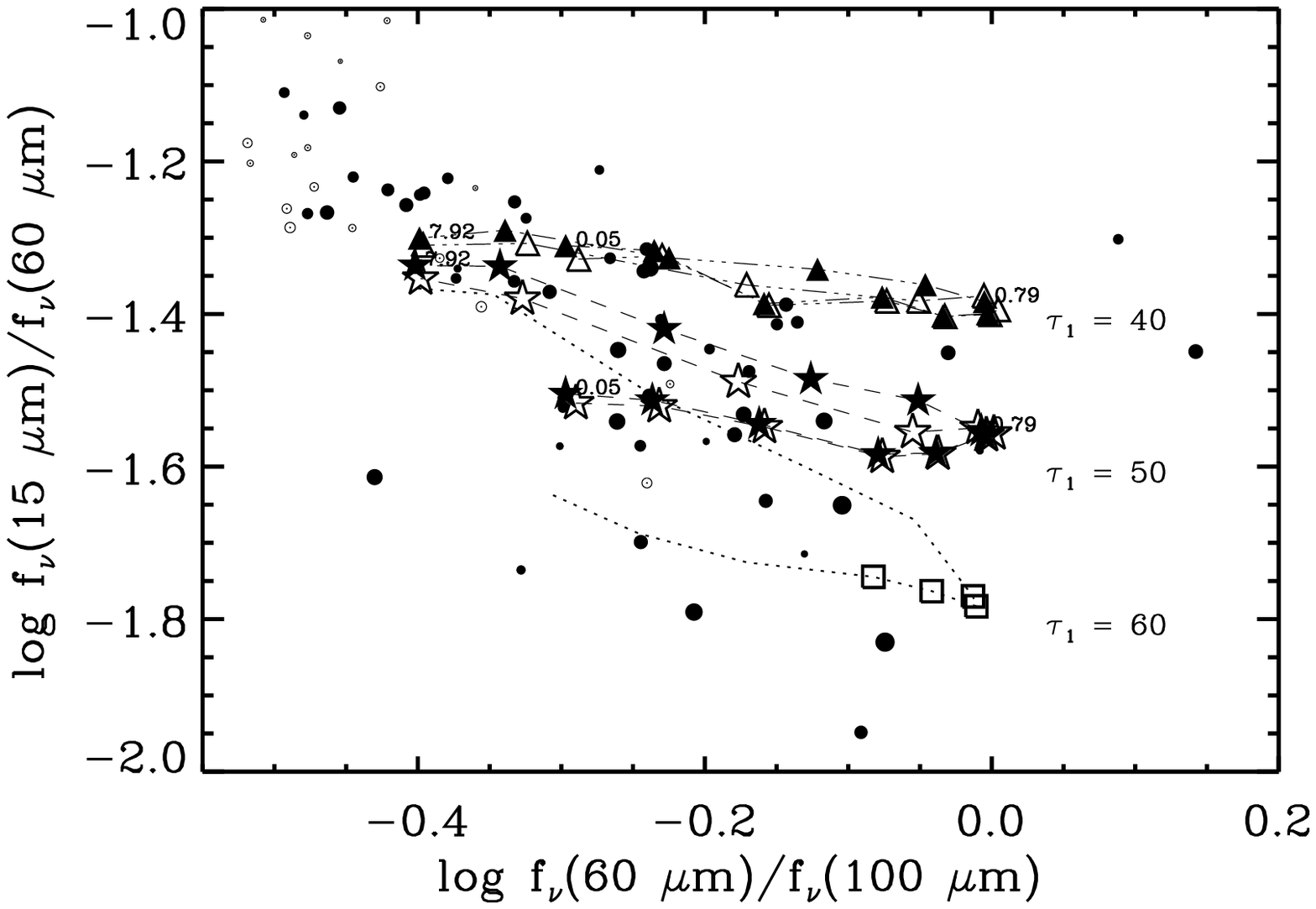}
\caption{IRAS-ISO color-color diagrams for the Dale et al. sample.
Black circles are the data. Different symbols are different
optical depths of MCs ($\tau_{1\mu m}$) and escape times of newly
born stars from MCs ($t_\rmn{esc}$). The parameters of the models
are $t_\rmn{b} =20$ Myr, fraction of molecular gas mass to
total gas mass=0.5. The ticks correspond to
$\frac{age_\rmn{b}}{t_\rmn{b}}=$ 0.05, 0.08, 0.12, 0.20, 0.31,
0.50, 0.80, 1.26, 2.00, 3.15, 5.0, and 7.92. The lines connect the
different evolution sequences. In order to remark the region
occupied by the normal spiral galaxies, we also plot the Boselli
et al. (2003) sample as open circles (see text for details). The
sizes of the symbols represent the FIR luminosities as marked by
the legend in the top right panel.}
 \label{fig:fin}
\end{figure*}

\subsection{Results of the comparison}

In the four panels of Fig. \ref{fig:fin} we plot the observed ISO LW2 and LW3 and IRAS 12, 25, 60
and 100 $\mu$m flux ratios. The LW2 filter includes the $6.2$, $7.7$, and $8.6 \mu$m PAH bands
while the LW3 filter includes the $11.9$, $12.7$ and $16.4$ bands, but it is less affected by PAHs
as compared to the LW2. Finally also the IRAS 12$\mu$m band is dominated by the emission of the
$11.3$, $11.9$ and $12.7$ PAHs.

Nearby star forming galaxies are known to follow a well-defined
trend in the IRAS flux density ratios $f_\nu (12 \mu \rmn{m})
/f_\nu (25 \mu \rmn{m})$ versus $f_\nu (60 \mu \rmn{m}) /f_\nu
(100 \mu \rmn{m})$ (Helou 1986; Dale et al. 2001). This trend is
in the sense that at increasing $f_\nu (60 \mu \rmn{m}) /f_\nu
(100 \mu \rmn{m})$ the $f_\nu (12 \mu \rmn{m}) /f_\nu (25 \mu
\rmn{m})$ and $f_\nu (6.75 \mu \rmn{m}) /f_\nu (15 \mu \rmn{m})$
ratios decrease. This has been explained as an increasing sequence
of star formation activity. The warmer $f_\nu (60 \mu \rmn{m})
/f_\nu (100 \mu \rmn{m})$ color reflects the fact that higher dust
temperatures of big grains (in thermal equilibrium) shift the FIR
emission peak to shorted wavelengths. The common interpretation of
the $f_\nu (12 \mu \rmn{m}) /f_\nu (25 \mu \rmn{m})$, $f_\nu (6.75
\mu \rmn{m}) /f_\nu (15 \mu \rmn{m})$ behavior is that, at the
same time, the stronger radiation field destroys very small grains
and PAHs (see Section~\ref{susec:mc})  and enhances the MIR
continuum as well.

In that figure, we have superimposed the time evolution of models
computed with three different values of the MCs optical depth,
$\tau_\rmn{1}$= 40, 50 and 60, and two escape timescales of newly
born stars from their parent MCs, $t_\rmn{esc}$=10 Myr and
$t_\rmn{esc}$=30 Myr. The time evolution is indicated by the ratio
between the age of the starburst and the e-folding time of the
star formation process ($age_\rmn{b}/t_\rmn{b}$), from the
beginning of the starburst phase ($age_\rmn{b}/t_\rmn{b} \sim
0.05$) to the quiescent phase ($age_\rmn{b}/t_\rmn{b} \sim 4.0 -
7.92$). The ticks correspond to different values of the ratio
$age_\rmn{b}/t_\rmn{b} =$ 0.05, 0.08, 0.12, 0.20, 0.31, 0.50,
0.80, 1.26, 2.00, 3.15, 5.0, and 7.92, and the evolution is
defined by the lines. The SEDs of our models are initially
dominated by emission of the underlying quiescent disk component
which is characterized by cold FIR colors (Log($f_\nu (60 \mu
\rmn{m}) /f_\nu (100 \mu \rmn{m})$) $\le$-0.3). As their
luminosity increases, due to the rapid formation of new massive
stars, they evolve toward hotter FIR colors, reaching a maximum
when the starburst is well developed (Log($f_\nu (60 \mu \rmn{m})
/f_\nu (100 \mu \rmn{m})) \ge$0). Then they turn back toward
cooler FIR colors (but with higher $f_\nu (12 \mu \rmn{m}) /f_\nu
(25 \mu \rmn{m})$ and $f_\nu (6.75 \mu \rmn{m}) /f_\nu (15 \mu
\rmn{m})$ colors) as the star formation declines and the oldest
stars leave the molecular clouds. This behavior is common to all
the selected starburst models.

The agreement with the data is fairly good. In addition we may
draw the following considerations.

We did not include a PAH destruction mechanism continuously
increasing with the intensity of the radiation field, instead we
made the assumption that within star forming regions PAHs are
damped by a fixed factor (see Sec. \ref{susec:mc}). Therefore the
lowering of the $f_\nu (6.75 \mu \rmn{m}) /f_\nu (25 \mu \rmn{m})$
and $f_\nu (6.75 \mu \rmn{m}) /f_\nu (15 \mu \rmn{m})$ colors in
the models simply reflects the progressive dominance of the MC
component over the cirrus one, when the starburst is in its star
formation peak. The sequence of colors is then interpreted as an
evolutionary sequence characterized by the dominance of the MC
over the cirrus component (Helou 1986; Helou et al. 1991). Four
main evolutionary phases can be recognized: the early starburst
phase characterized by cool $f_\nu (60 \mu \rmn{m}) /f_\nu (100
\mu \rmn{m})$ colors and low MIR-FIR colors; the peak starburst
phase with the models reaching the hottest FIR colors and the
lowest MIR-FIR colors; the evolved starburst phase with warm FIR
colors and warm MIR-FIR colors; finally the post starburst phase
where the models move toward the cool FIR colors and hot MIR-FIR
colors, typical of normal spiral galaxies.

The models considered here do not reach the coolest FIR region of
the diagram ($f_\nu (60 \mu \rmn{m}) /f_\nu (100 \mu \rmn{m}) \leq
-0.45$). This is the region dominated by normal spiral galaxies as
can be seen by the location of the Boselli et al. (2003) Virgo
spiral galaxy sample, plotted for the purpose of comparison (open
circles). In this paper we have considered models meant to
reproduce actively star forming galaxies. It is conceivable that
the coolest region requires a more direct consideration of the
PDRs (see Section~\ref{susec:mc}).

The evolution of the starbursts in the diagrams is modulated by
the optical depth of the molecular clouds, which affects primarily
the colors $f_\nu (6.75 \mu \rmn{m}) /f_\nu (60 \mu \rmn{m})$ and
$f_\nu (15 \mu \rmn{m}) /f_\nu (60 \mu \rmn{m})$. At larger optical
depths these flux ratios decrease.  The flux ratio $f_\nu (6.75
\mu \rmn{m}) /f_\nu (15 \mu \rmn{m})$ is instead less dependent on
$\tau_1$, and in this case both data and models show a lower
dispersion. Furthermore the range of allowed optical depths is
fairly narrow, ranging from $\tau_1\sim$40 to $\sim$60. Only a few
objects would require a slightly higher optical depth. These
objects are particular in the sense that they do not share the
trend common to all the others and the majority of them are
actually ULIRGs, which could be highly obscured (Prouton et al.
2004).

The effect of other tested parameters are in brief the following: shorter e-folding times
$t_\rmn{b}$ of the burst produce hotter peak models. However these models are also characterized by
a wider evolutionary sequence in the MIR colors; in practice with $t_\rmn{b} \lesssim 15$ Myr, the
evolutionary path is an envelope of the data without matching them. For values of $t_\rmn{b}
\gtrsim 25$ Myr, the evolution of the MIR colors is as observed but the range of the $60/100$ color
is very small and concentrated to the warm central region. The escape timescale of young stars from
their parent MCs is well constrained by this comparison. Short escape times, $t_\rmn{esc}\lsim 10$
Myr, result in too wide evolutionary sequences, in the MIR-FIR plots, again surrounding the data,
while values higher than $t_\rmn{esc}\sim 35$ Myr are confined to low MIR-FIR flux ratios. The
typical escape time suggested by the data is of $\sim 20$ Myr, comparable to the e-folding time of
the SFR. As for the molecular gas fraction, the comparison suggests that it should be confined
between 30\% and 70\% of the total gas mass, with 50\% the best value.

Summarizing, from the comparison with the data we find the
following typical parameters for our models: a star formation
e-folding time $t_\rmn{b} \sim 15 - 25$ Myr; fractions of
molecular mass between 0.3 to 0.7, with the hottest models having
the lowest molecular gas fraction; escape times $t_\rmn{esc}$
between 10 to 40 Myr.

\subsection{ISO-IRAS-Radio color-color plots}

\begin{figure*}
\centering
\includegraphics[angle=0,width=8.8truecm]{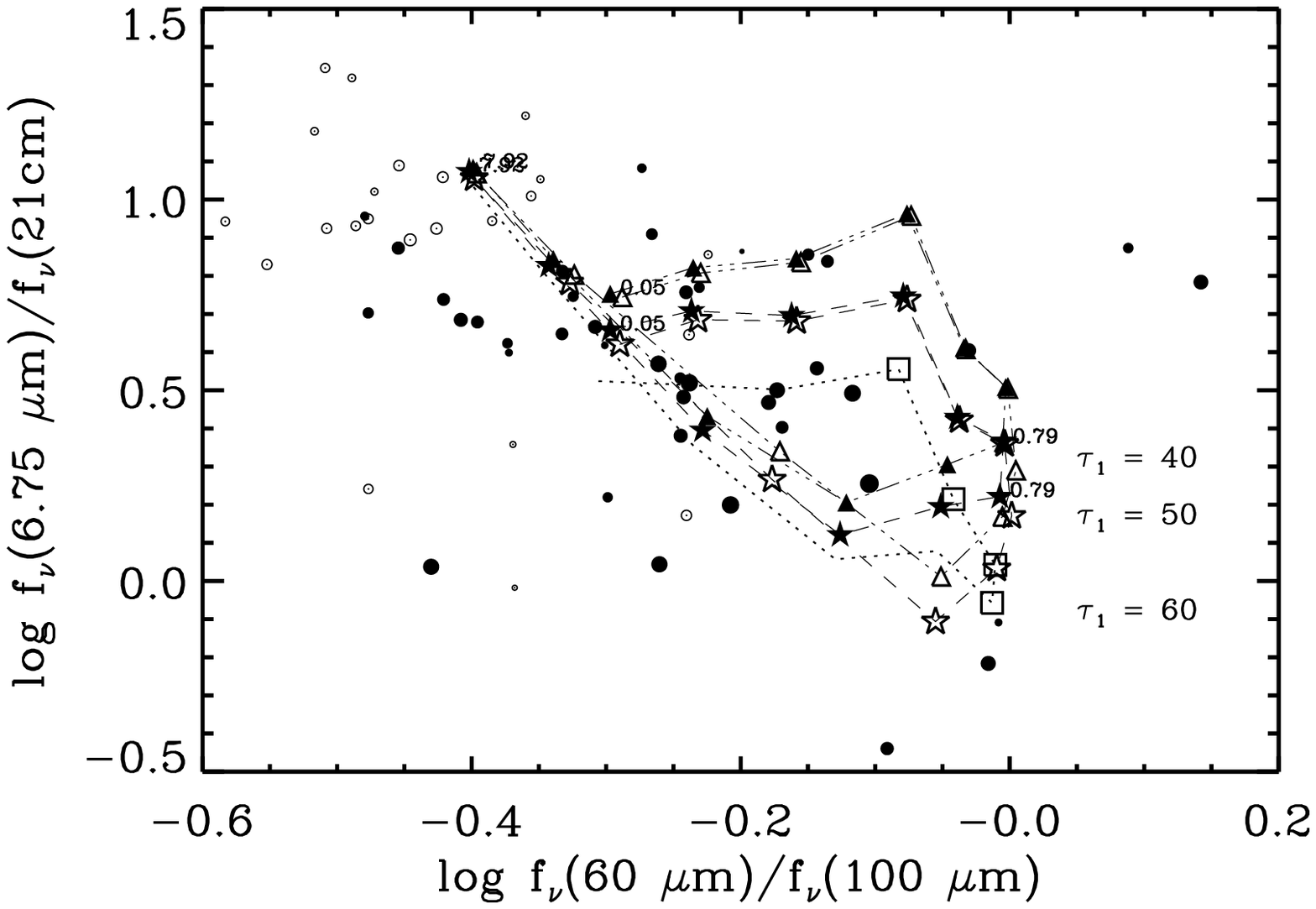}
\includegraphics[angle=0,width=8.8truecm]{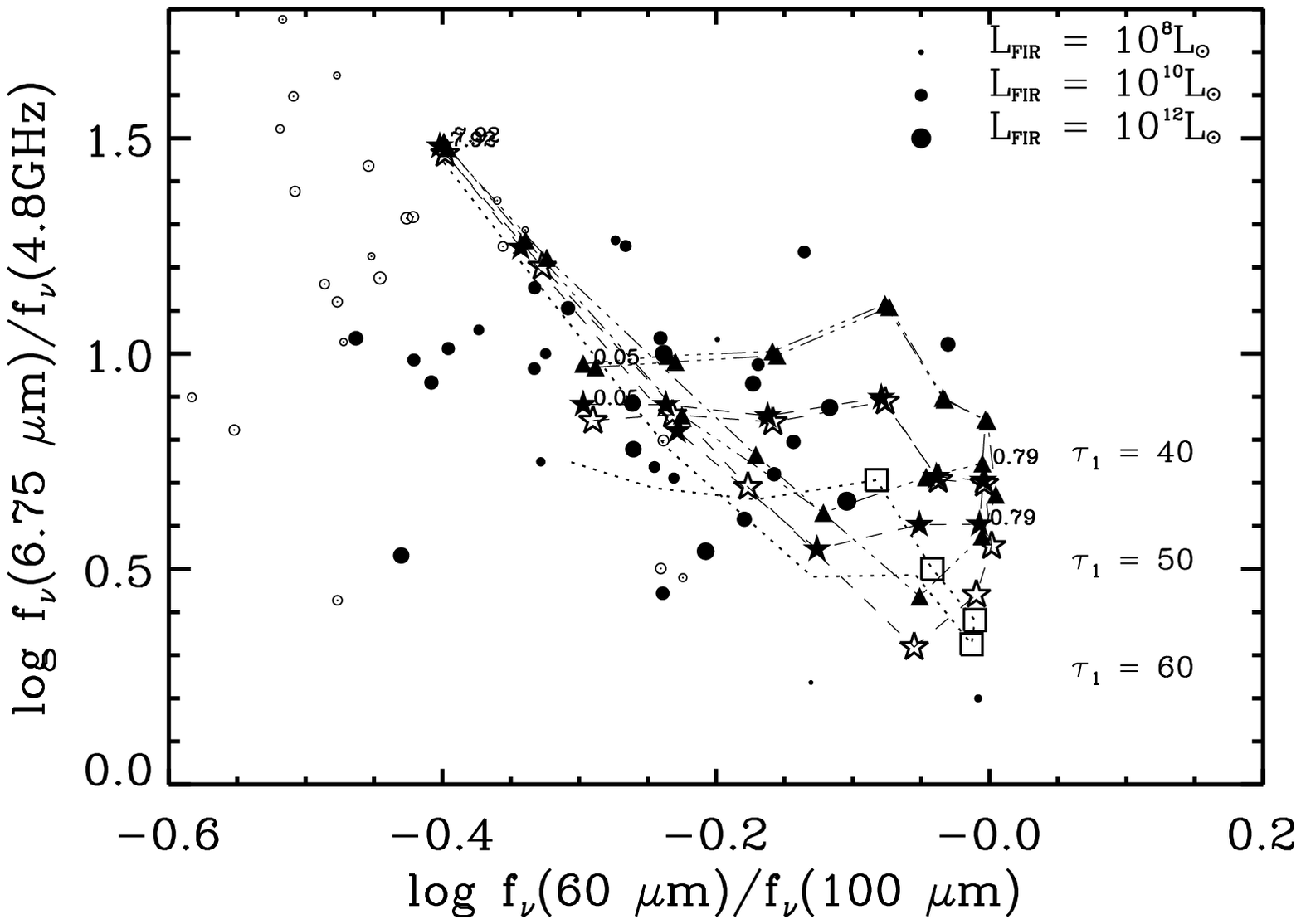}
\includegraphics[angle=0,width=8.8truecm]{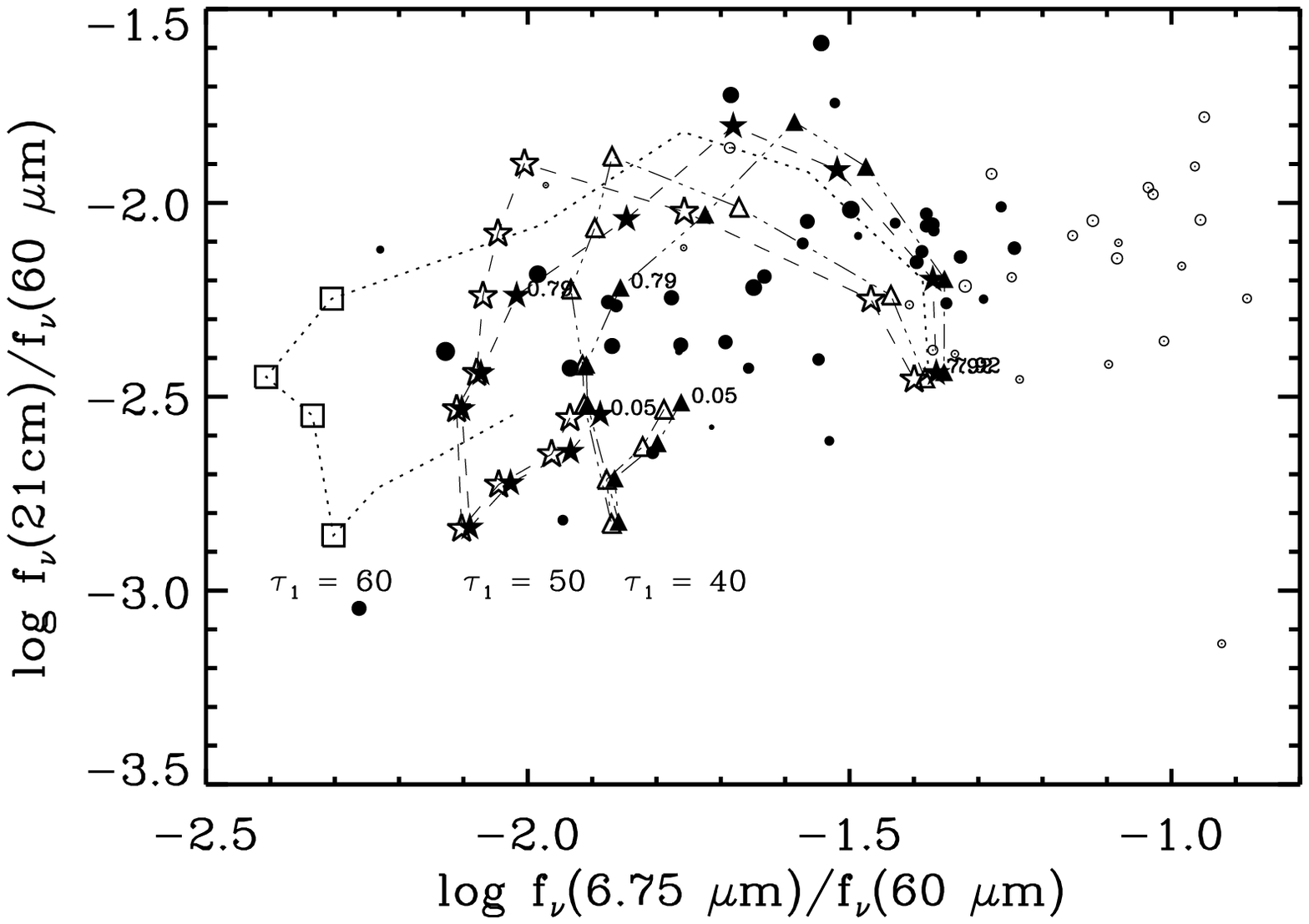}
\includegraphics[angle=0,width=8.8truecm]{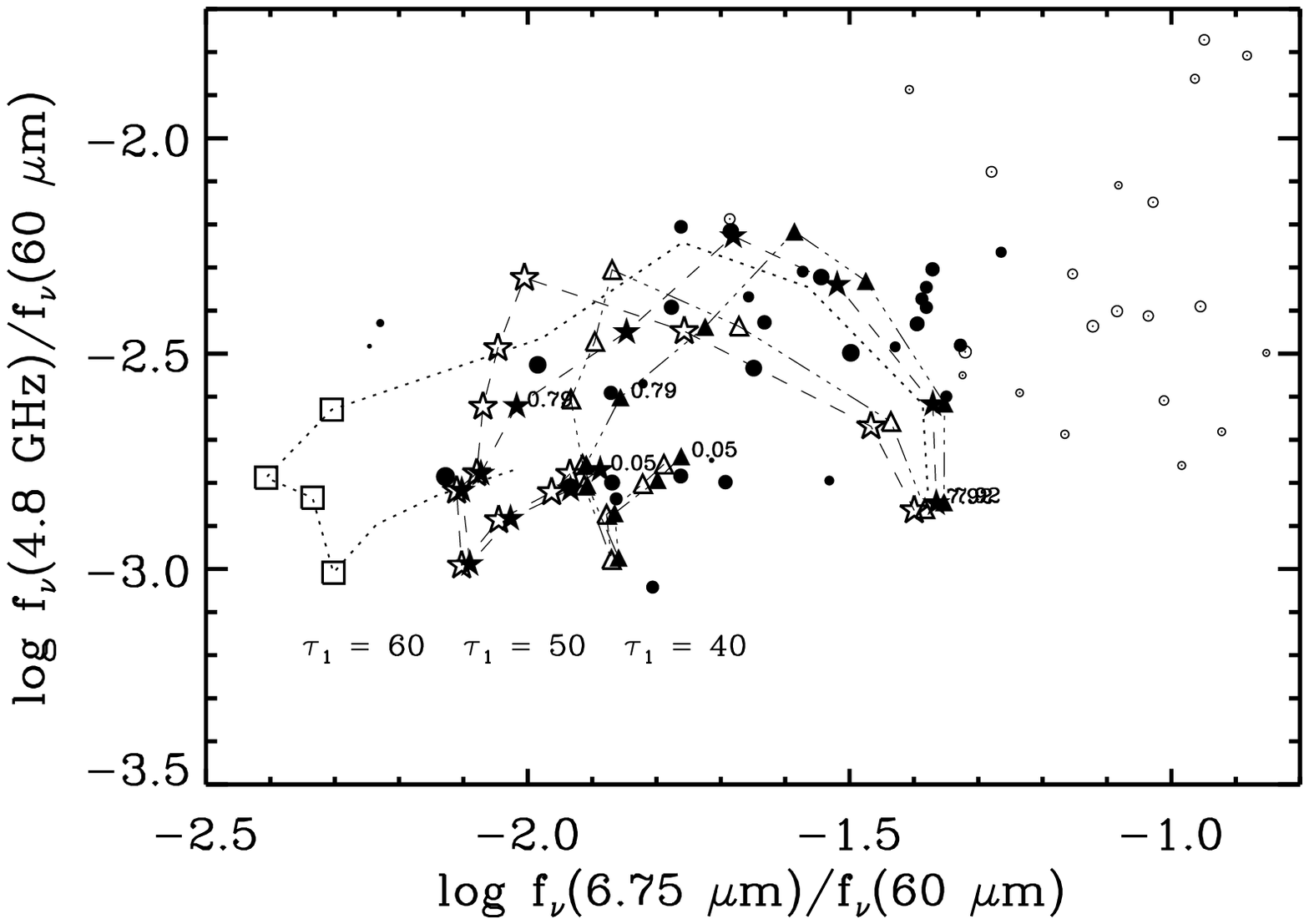}
\caption{IRAS-ISO-radio color-color diagrams for the Dale et al.
sample. Black circles are the data. Different symbols are
different optical depths $\tau_1$ and escape timescales
$t_\rmn{esc}$. The parameters of the models are: $t_\rmn{b} =20$
Myr, fraction of molecular mass = 0.5. The ticks correspond
to $\frac{age_\rmn{b}}{t_\rmn{b}}=$ 0.05, 0.08, 0.12, 0.20, 0.31,
0.50, 0.80, 1.26, 2.00, 3.15, 5.0, and 7.92. The lines connect the
different evolutive sequences. In order to remark the region
occupied by the normal spiral galaxies, we plot the Boselli et al.
(2003) spiral galaxy sample as open circles. The sizes of the
symbols represents the FIR luminosities.} \label{fig:rad}
\end{figure*}

For the majority of the galaxies radio data at 6.2 and 21cm are available (see Table
\ref{tab:tabdale}). The radio emission of the models is computed following the prescriptions given
in Bressan, Silva \& Granato (2002). Thermal emission is computed according to the flux of ionizing
photons, while the non thermal emission is set proportional to the core collapse supernova rate.

The FIR to radio ratio $q$ is defined as:

\begin{equation}
q=\log \frac{F_\rmn{FIR}/(3.75\times 10^{12}\mbox{Hz})}{f_\nu
(1.49\mbox{GHz})/(\mbox{W m}^{-2}\mbox{Hz}^{-1})}\simeq 2.35\pm
0.2 \label{qeq}
\end{equation}

where $F_\rmn{FIR}=1.26 \, 10^{-14}(2.58\, f_\nu(60\mu{\rm m}) +
f_\nu(100\mu{\rm m}))$ W~m$^{-2}$, with $f_\nu(60 \mu \rm m)$
and $f_\nu(100 \mu \rm m)$ in Jy. The average observed value of the q
parameter for a spiral galaxy is $q \simeq 2.35$ at 1.49 GHz
(Condon 1992; Sanders \& Mirabel 1996)

Fig. \ref{fig:rad} shows the evolution of the models in composite MIR-FIR-radio bands. The
agreement between models and data is fairly good, consistent with the fact that the models well
reproduce the observed FIR-radio correlation. The partial disagreements in the 21 cm plots for
models during the peak phase may be explained by the need to account for free-free absorption
(Bressan et al. 2002; Prouton et al. 2004). Again, the region of very quiescent spiral galaxies is
not covered by the starburst models.

\section{Conclusions}
\label{sec:conc}

This paper is devoted to an accurate analysis of the properties of starburst galaxies in the mid
infrared. This spectral region is dominated by PAH emission features, and an update with respect to
our previous treatment of PAHs in our code GRASIL (Silva et al. 1998) was necessary in its two main
components, the cirrus and the molecular clouds, respectively.

In the cirrus component we have introduced
the new quantitative description of PAHs in the
diffuse ISM, given by Li \& Draine (2001).
As for the MC component, the observational evidence of a lack of PAH
emission in regions characterized by strong UV radiation fields,
forced us to reduce the PAHs abundance
by a large factor (i.e. $\sim 1000$ times less than in the cirrus component).

As a workbench we have compared the updated model with the SEDs of a select sample  of star forming
galaxies, for which both accurate MIR narrow band and NIR, FIR and Radio observations are
available. This comparison probes successfully over the wide range of FIR luminosities of the
selected galaxies.

We have then analyzed the observed trend of MIR-FIR and radio colors for a larger sample of
galaxies, from actively star forming to obscured starbursts. To this aim we have compared a large
library of model SEDs, spanning a wide range of physical parameters, with observed color-color
plots. The data are well reproduced by the models within a limited range of values of the main
physical parameters.

As already suggested by Helou (1986) and Dale et al (2001), the observed MIR-FIR trend can be
interpreted as a sequence of the dominance of the MC component with respect to the cirrus
component. However, contrary to previous works,  that were   based on mainly empirical SEDs, this
comparison allows us to propose that the observed sequence corresponds to the evolution of the
starburst phenomenon.

\begin{figure*}
\centering
\includegraphics[angle=0,width=12truecm]{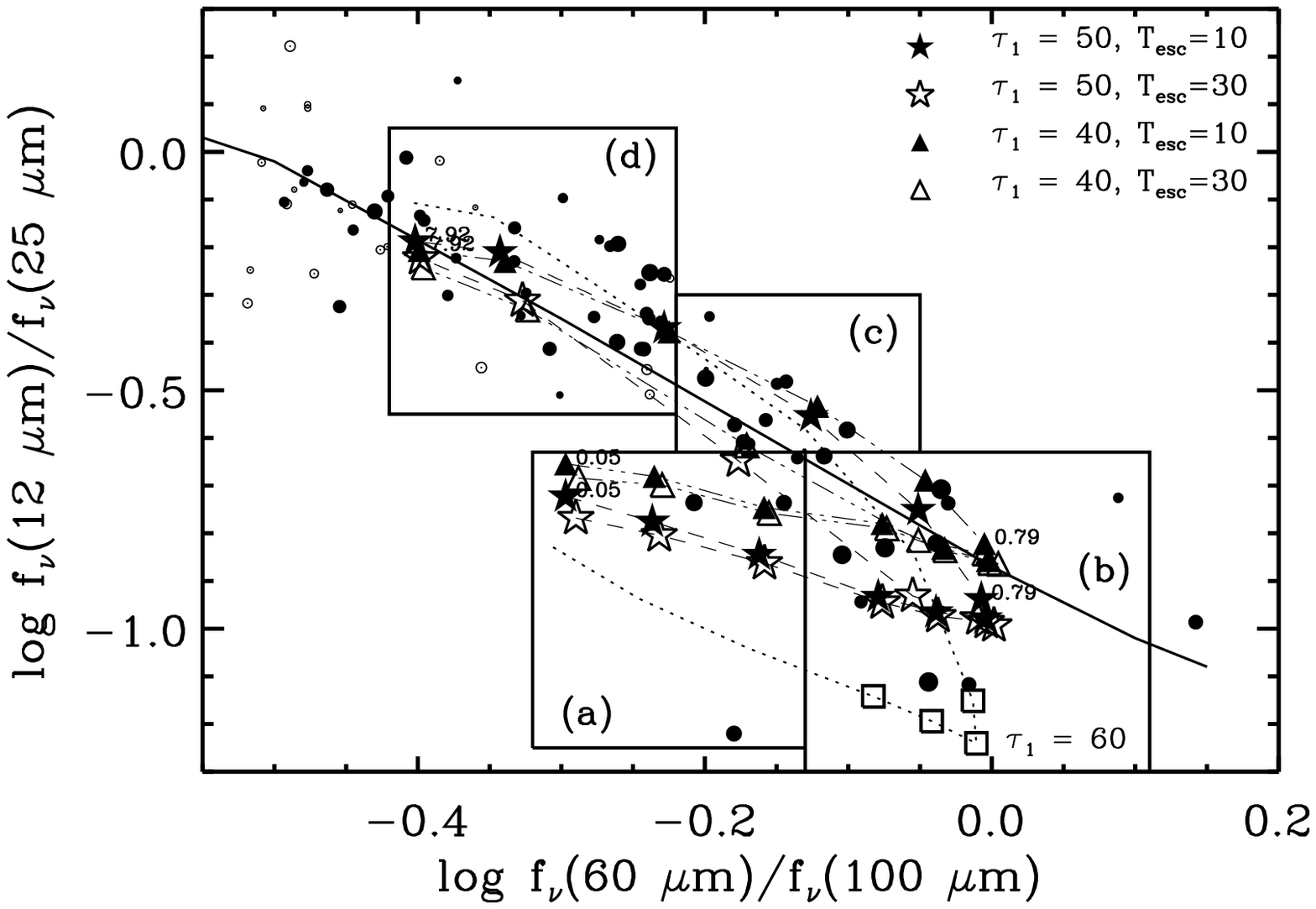}
\caption{IRAS color-color diagrams for the Dale et al. sample. Meaning of symbols is as in Fig.
\ref{fig:fin}. The four closed regions highlight the four main evolutionary phases of our starburst
models with which we interpret the behavior of data. See text for more details.} \label{fig:coclu}
\end{figure*}

Four main phases can be highlighted, as summarized in Fig. \ref{fig:coclu}:

\begin{itemize}
\item[(a)] The early starburst phase,  characterized by warm FIR and cold MIR colors and where the
few newly formed stars are still deep inside the progenitor molecular clouds. The SED is dominated
by emission from MCs but the intensity of the radiation field has not reached its peak value. In
our model this phase has a low ratio of $age_\rmn{b}/t_\rmn{b}$, between 0.05 to 0.20.

\item[(b)] The starburst peak phase, where the majority of massive stars are produced, and the SED
is dominated by the hot dust emission from the molecular cloud component. During this phase the
starburst reaches the hottest FIR colors and the lowest relative PAH emission. It is characterized
by $age_\rmn{b}/t_\rmn{b}$ between 0.30 to 2.0, and the young stars are still mostly embedded in the parent
molecular clouds.

\item[(c)] The evolved starburst phase, where the number of newly formed stars has drastically
decreased and, at the same time, most of the still hot stars are out of the progenitor MCs and heat
the cirrus. The emission in this phase, which lasts between $age_\rmn{b}/t_\rmn{b}\simeq$  2.0 to 4.0, is
dominated by the cirrus component.

\item[(d)] The post-starburst phase, $age_\rmn{b}/t_\rmn{b} > 4$, where the current SFR is mainly due to
the quiescent disk and all the stars formed during the starburst
are now outside the MC. The FIR and MIR colors are evolving toward
those of normal spiral galaxies.
\end{itemize}

The data suggest that the starburst phenomenon is characterized by
a surprising homogeneity. The optical depth of the molecular
clouds at 1 $\mu$m is confined between 40 and 60; the escape time
scale from molecular clouds is between 10 and 30 Myr and it
mainly affects the advanced evolution of the starburst; the
e-folding time of the star formation process cannot be shorter
than about 15 Myr, nor larger than about 30 Myr. Though
those figures are similar to the estimates given in Silva et al.
(1998), only with the advent of accurate measurements in the mid
infrared it has been possible to restrict them to a very narrow
range.

It is also evident from the models, that PAH features cannot be taken as a direct measure of the
dominant powering mechanism since, in those cases where the MC component dominates, there is a
significant lack of emission, as observed in AGNs.

Further work is needed to reproduce the most quiescent galaxies
(occupying the region with the coldest FIR colors, Log($f_\nu (60
\mu \rmn{m}) /f_\nu (100 \mu \rmn{m})$) $\le -0.45$), where possibly
a more complex description and modelling of the different zones of
star formation, for instance the PDR zones, is required, and will
be the subject of a forthcoming paper.

\section*{Acknowledgements}
We would like to thank the anonymous referee whose comments have helped to improve the paper.
O. Vega  and M. Chavez acknowledge the support of INAOE and the
Mexican CONACYT project  36547 E; O. Vega  also acknowledges the
support of the Mexican CONACYT project  39714 F; L. Silva, A.
Bressan and G. L. Granato acknowledge warm hospitality by INAOE.

 \label{lastpage}


\begin{thebibliography}{99}

\bibitem[\protect\citeauthoryear{Abergel et al.}{1994}]{aber94}
Abergel A., Boulanger F., Mizuno A., Fukui Y., 1994, ApJ, 423,~59

\bibitem[\protect\citeauthoryear{Allain et al.}{1996}]{alla96}
Allain T., Leach S., Sedlmayr E., 1996, A\&A, 305, 616

\bibitem[\protect\citeauthoryear{Allamandola et al.}{1985}]{alla85}
Allamandola L. J., Tielens A. G. G. M., Barker J. R., 1985, ApJ, 290, 25

\bibitem[]{} Armus, L. et al., 2004, ApJS, 154, 178

\bibitem[\protect\citeauthoryear{Beintema et al.}{1996}]{beinte96}
Beintema D. A., van den Ancker M. E., Molster F. J.,  Waters
L. B. F. M., Tielens A. G. G. M., Waelkens C., de Jong T., de
Graauw T., Justtanont K., Yamamura I., Heras A., Lahuis F.,
Salama A., 1996, A\&A, 315, 369

\bibitem[]{} Bendo et al., 2002, AJ, 123, 3067

\bibitem[\protect\citeauthoryear{Boselli et al.}{2003}]{bose03}
Boselli A., Sauvage M., Lequeux J., Donati A., Gavazzi G., 2003,
A\&A, 406, 867

\bibitem[\protect\citeauthoryear{Boselli et al.}{2004}]{bose04}
Boselli A., Lequeux J., Gavazzi G., 2004, A\&A, 428, 409

\bibitem[\protect\citeauthoryear{Boulanger et al.}{1996}]{boul96}
Boulanger F. et al., 1996, A\&A, 315, 325

\bibitem[\protect\citeauthoryear{Boulanger}{2000}]{boul00}
Boulanger F., 2000, ISO Beyond Point Sources: Studies of Extended Infrared Emission, ESA-SP 455, p. 3

\bibitem[\protect\citeauthoryear{Bressan et al.}{2002}]{bres02}
Bressan A., Silva L., Granato G. L., 2002, A\&A, 392, 377

\bibitem[]{} Carico D. P., Keene J., Soifer B. T., Neugebauer
G., 1992, PASP, 104, 1086

\bibitem[\protect\citeauthoryear{Cesarsky et al.}{1996}]{cesa96}
Cesarsky C. J., et al., 1996, A\&A, 315, 32

\bibitem[]{} Chini R., Kruegel E., Lemke R., Ward-Thompson D.,
1995, A\&A, 295, 317

\bibitem[]{} Chini R., Kruegel E., Lemke R., 1996, A\&AS, 118,
47

\bibitem[\protect\citeauthoryear{Clavel et al.}{2000}]{clav00}
Clavel J., Schulz B., Altieri B., Barr P., Claes P., Heras A.,
Leech K., Metcalfe L.,  Salama, A.,  2000, A\&A, 357, 839

\bibitem[\protect\citeauthoryear{Condon et al.}{1990}]{condon90}
Condon J. J., Helou G., Sanders D. B., \& Soifer B. T., 1990,
ApJS, 73, 359

\bibitem[\protect\citeauthoryear{Condon et al.}{1991}]{condon91}
Condon J. J., Frayer D. T., Broderick, J. J., 1991, AJ, 101,
362

\bibitem[\protect\citeauthoryear{Condon}{1992}]{condon92}
Condon J. J., 1992, ARA\&A, 30, 575

\bibitem[\protect\citeauthoryear{Condon et al.}{1995}]{condon95}
Condon J. J., Frayer D. T., Broderick J. J., 1995, ADIL, JC,
02

\bibitem[\protect\citeauthoryear{Contursi et al.}{2000}]{cont00}
Contursi A., Lequeux J., Cesarsky D., Boulanger F., Rubio M.,
Hanus M., Sauvage M., Tran D., Bosma A., Madden S., Vigroux L.,
2000, A\&A, 362, 310

\bibitem[\protect\citeauthoryear{Dale et al.}{2000}]{dale00}
Dale D. et al., 2000, ApJ, 120, 583

\bibitem[]{} Dale D., Helou G., Contursi A., Silbermann N.
A., Kolhatkar S., 2001, ApJ, 459, 215

\bibitem[]{} Dale D. et al., 2005, astro-ph/0507645

\bibitem[]{} Dopita M. A., Groves B. A., Fischera J., Sutherland
R. S., Tuffs R. J., Popescu C. C., Kewley L. J., Reuland M.,
Leitherer C., 2005, ApJ, 619, 755

\bibitem[]{} Draine B. T., Lee H. M., 1984, ApJ, 285, 89

\bibitem[]{} Draine B. T., 2003, ARA\&A, 41, 241

\bibitem[]{} Dunne L., Eales S., Edmunds M., Ivison R., Alexander P., Clements D.
L., 2000, MNRAS, 315, 115

\bibitem[]{} Dwek E., 2004, ApJ, 611, L109

\bibitem[]{} Engelbracht C. W., Gordon K. D., Rieke G. H., Werner M.
W., Dale D. A., Latter W. B., 2005, ApJ, 628, 29

\bibitem[]{} Forster Schreiber N. M., Sauvage M., Charmandaris V., Laurent O., Gallais P.,
Mirabel I. F., Vigroux L., 2003, A\&A, 399, 833

\bibitem[]{} Forster Schreiber, N. M., Roussel H., Sauvage M.,  Charmandaris V., 2004, A\&A,
419, 501

\bibitem[]{} Galliano F., Madden S. C., Jones A. P., Wilson C. D., Bernard,
J.-P.,  Le Peintre F., 2003, A\&A, 407, 159

\bibitem[]{} Galliano F., Madden S. C., Jones A. P., Wilson C. D., Bernard,
J.-P., 2005, A\&A, 434, 867

\bibitem[]{} Genzel R. et al., 1998, ApJ, 498, 579

\bibitem[]{} Granato G. L., Lacey C. G., Silva L., Bressan A., Baugh C.
M., Cole S.,   Frenk C. S., 2000, ApJ, 542, 710

\bibitem[]{} Helou G., 1986, ApJ, 311, L33

\bibitem[]{} Helou G., Ryter C.,  Soifer B. T., 1991, ApJ, 376, 505

\bibitem[]{} Helou G. et al., 1996, AAS, 189, 6704

\bibitem[]{} Helou G., Lu N. Y., Werner M. W., Malhotra S.,  Silbermann
N., 2000, ApJ, 532, 21

\bibitem[]{} Hony S., Van Kerckhoven C., Peeters E., Tielens
A. G. G. M., Hudgins D. M.  Allamandola L. J., 2001, A\&A, 370,
1030

\bibitem[]{} Israel F. P., van der Hulst J. M., 1983, AJ, 88,
1736

\bibitem[]{} Laor A.,  Draine B. T., 1993, ApJ, 402, 441

\bibitem[]{} Laurent O., Mirabel I. F., Charmandaris V., Gallais
P., Madden S. C., Sauvage M., Vigroux L., Cesarsky C., 2000, A\&A,
359,~887

\bibitem[]{} Leger A.,  Puget J. L., 1984, A\&A, 137, 5

\bibitem[]{} Leger A., D'Hendecourt L.,  Defourneau D., 1989, A\&A, 216, 148

\bibitem[]{} Li A.,  Draine B. T., 2001, ApJ, 554, 778

\bibitem[]{} Lu N., Helou G., Werner M. W., Dinerstein H. L.,
Dale D. A., Silbermann N. A., Malhorta S., Beichman C. A.,
Jarret T. H., 2003, ApJ, 588, 199

\bibitem[]{} Lutz D., Genzel R., Kunze D., Spoon H. W. W., Sturm
E., Sternberg A.,  Moorwood A. F. M., 1998, ASPC, 132, 89

\bibitem[]{} Machalski J., Condon J. J., 1999, ApJS, 123, 41

\bibitem[]{} Madden S., 2000, NewAR, 44, 249

\bibitem[]{} Madden S., 2005, AIPC, 761, 223

\bibitem[]{} Marty P., Serra G., Chaudret B., Ristorcelli
I., 1994, A\&A, 282,~916

\bibitem[]{} Mattila K., Lemke D., Haikala L. K., Laureijs R. J., Leger
A., Lehtinen K., Leinert C.,  Mezger P. G., 1996, A\&A, 315,
353

\bibitem[]{} Metcalfe L., Steel S. J., Barr P., Clavel J., Delaney M.,
Gallais P., Laureijs R. J., Leech K.,  McBreen B., Ott S.,
Smith N., Hanlon L., 1996, A\&A, 315, 105

\bibitem[]{} Moutou C., Leger A., D'Hendecourt L., 1996, A\&A, 310, 297

\bibitem[]{} Niklas S., Klein U., Braine J., Wielebinski R., 1995, A\&AS, 114,~21

\bibitem[]{} Panuzzo P., Bressan A., Granato G. L., Silva L.,
 Danese L., 2003, 409, 99

\bibitem[]{} Panuzzo P., Silva L., Granato G. L., Bressan A., Vega O.,
2005, in Popescu C. C., Tuffs R. J., eds, AIP Conference Proceedings 761, p. 187

\bibitem[]{} Peeters E., Spoon H. W. W.,  Tielens A. G. G. M., 2004a, ApJ, 613, 986

\bibitem[]{} Peeters E., Allamandola L. J., Hudgins D. M., Hony S.,  Tielens A. G. G.
M., 2004b, in Witt N. A., Clayton G. C., Draine B. T., eds, ASP
Conference Series, Vol. 309, p.141

\bibitem[]{} Peeters E., Mattioda A. L., Hudgins D. M., Allamandola L.
J., 2004c, ApJ, 617, 65

\bibitem[]{} Prouton O. R., Bressan A., Clemens M., Franceschini A., Granato G. L.,  Silva
L., 2004, A\&A, 421, 115

\bibitem[]{} Puget J. L.,  Leger A., 1989, ARA\&A, 27, 161

\bibitem[]{} Rigopoulou D., Spoon H. W. W., Genzel R., Lutz D., Moorwood
A. F. M.,  Tran Q. D., 1999, AJ, 118, 2625

\bibitem[]{} Roussel H., Sauvage M., Vigroux L., Bosma A., 2001, A\&A, 372,~427

\bibitem[]{} Salpeter E. E., 1955, ApJ, 121, 161

\bibitem[]{} Sanders D. B.,  Mirabel I. F., 1996, ARA\&A, 34, 749

\bibitem[]{} Serra G., Chaudret B., Saillard Y., Le Beuze A., Rabaa
H., Ristorcelli I.,  Klotz A., 1992, A\&A, 260, 489

\bibitem[]{} Silva L., Granato G. L., Bressan A.,  Danese L.,
1998, ApJ, 509,~103

\bibitem[]{} Silva L., 1999, PhD Thesis, SISSA, Trieste, Italy

\bibitem[]{} Smith J. D. T. et al., 2004, ApJS, 154, 199

\bibitem[]{} Tacconi-Garman L. E., Sturm E., Lehnert M., Lutz D., Davies
R. I.,  Moorwood A. F. M., 2005, A\&A, 432, 91

\bibitem[]{} Tielens A. G. G. M., Hony S., van Kerckhoven C.,  Peeters
E., 1999, in The Universe as Seen by ISO. Eds. P. Cox  M. F.
Kessler. ESA-SP 427., p. 579

\bibitem[]{} Tielens A. G. G. M., 1999, Formation and Evolution of Solids in Space, Edited by J.
Mayo Greenberg  Aigen Li. Kluwer Academic Publishers, p.33

\bibitem[]{} Verstraete L., Puget J. L., Falgarone E., Drapatz S., Wright C. M.,
 Timmermann, R., 1996, A\&A, 315, 337

\bibitem[]{} Vigroux L., Mirabel F., Altieri B., Boulanger F., Cesarsky
C., Cesarsky D., Claret A., Fransson C., Gallais P., Levine
D., Madden S., Okumura K.,  Tran D., 1996, A\&A, 315, 93

\bibitem[]{} Weedman D. W., Hao L., Higdon S. J. U., Devost D., Wu Y., Charmandaris V., Brandl B., Bass E.,
 Houck J. R., 2005,  astro-ph/0507423


\bibitem[]{} Xu C., de Zotti G., 1989, A\&A, 225, 12

\bibitem[]{} Zubko V., Dwek E., Arendt R. G., 2004, ApJS, 152, 211

\end{thebibliography}
\end{document}